\documentclass[final,twocolumn,pra,aps,superscriptaddress]{revtex4-2}
\usepackage{graphicx,tabularx,epstopdf,amsmath}

\begin{document}
\title{ Modeled  vortex dynamics on a Bose-Einstein condensate  in a rotating  lattice}
\author{ P. Capuzzi}
\affiliation{Universidad de Buenos Aires, Facultad de Ciencias Exactas y Naturales,
Departamento de Física. Buenos Aires, Argentina.}
\affiliation{CONICET - Universidad de Buenos Aires, Instituto de Física de Buenos
Aires (IFIBA). Buenos Aires, Argentina.}
\author{D. M. Jezek}
\affiliation{CONICET - Universidad de Buenos Aires, Instituto de Física de Buenos
Aires (IFIBA). Buenos Aires, Argentina.}
\date{\today}
\begin{abstract}
  We study the dynamics of vortices in a Bose-Einstein condensate
  within a rotating four-site lattice which can be effectively
  described by a multimode model.  Such a vortex dynamics develops
  along the low-density paths that separate the sites, and it is
  ruled by the phase differences between them. Hence, by appropriately
  selecting the initial conditions for on-site populations and phase
  differences, one can access distinct types of evolutions.  We show
  that, by choosing equal populations in alternate sites, one can
  construct two-mode model Hamiltonians which allows us to model a
  large variety of associated vortex orbits. In particular, one can
  select the type of trajectory of the vortex and predict the creation
  and annihilation of vortex-antivortex pairs near the trap center.
  Estimates for the periods of closed vortex orbits and for the times
  that the vortices spend inside the lattice when dealing with open
  orbits, are obtained in terms of the two-mode models parameters and
  the rotation frequency only. We believe that the present study
  establishes a suitable platform to engineer different vortex
  dynamics.
\end{abstract}
\maketitle
\clearpage

\section{ Introduction}

The study of quantized vortices has been established as a central
topic in the research on liquid Helium for several decades, since they
represent a clear hallmark of superfluidity \cite{don91}.  Such an
interest has been latterly transferred to both experimental and
theoretical works in cold atom systems, and in particular in Bose
Einstein condensates (BECs)\cite{dal99}, in which the study of
different phenomena involving vortices has constituted a very active
area of research \cite{fet09}.

From the experimental point of view, such studies started with the
first successful creation and observation of a vortex in 1999
\cite{mat99}. This was followed by the measurement of the precession
frequency of a vortex moving around the axis of a harmonic trapping
potential \cite{brian00}.  Almost simultaneously, experiments on
producing arrays of many vortices were carried out with rotating
harmonic traps \cite{mad00,abo01,enge03} and then, quadratic plus
quartic potentials were employed to reach higher frequencies
\cite{bre04,sto05}.  Later on, experiments dealing with quantized
circulation have also included toroidal trapping potentials
\cite{ryu07,ec14,wil22}.  Finally, in rotating optical lattices,
different configurations of vortices were observed in triangular and
square lattices by Tung \textit{et al.}  \cite{tun06}. In particular,
for sufficiently large rotation intensity, they reported a structural
crossover to a vortex lattice. In addition, for a square lattice, the
number of nucleated vortices as a function of the rotation frequency
has been analyzed in Ref. \cite{wi10}.

Theoretical studies involving vortices also include different types of
confining potentials and atomic species.  Stationary configurations of
arrays of vortices have been studied in rotating quadratic plus
quartic potentials \cite{fet05,kim05} and in a nonrotating one
\cite{jez08}.  The theoretical study of vortex dynamics in
inhomogeneous nonrotating systems has constituted a challenging task
due to the diverse density profiles produced in the different types of
trapping potentials.  The first studies were devoted to the simpler
case where the traps have an axial symmetry, and hence a single vortex
precedes around such a symmetry axis.  More recently, in more general
inhomogeneous media, the vortex velocity has been derived in terms of
the local density and its derivatives
\cite{sheehy04,nil06,jez108,jez08,gros18}.  When several vortices are
present, the velocity of a vortex is also affected by the velocity
field generated by the rest of the vortices.  Such an effect has been
considered in a number of studies involving a few corotating vortices
\cite{nav13}, vortices in rotating BECs that describe chaotic dynamics
\cite{zhang19}, the dynamics of vortex dipoles \cite{frei10} observed
in Ref. \cite{nee10}, quantum turbulence \cite{nee13}, interacting
vortex lines in elongated BECs \cite{sera15}, doubly quantized
vortices \cite{ma06}, and three-vortex configurations \cite{se10}.

When the density itself varies as a function of time, the vortex
dynamics becomes still much more difficult to describe.  In the last
years, in double-well systems, the passage of vortices along the
junctions has been related to phase slips \cite{abad15}, and also the
movement of vortices has been observed in solitonic systems
\cite{don14,jez16}. However, in the first case such a vortex passage
has been difficult to trace because it occurs during much shorter
timescales than the dynamics of the involved macroscopic variables,
and also because vortex-antivortex pairs were generated or annihilated
spontaneously during the evolution. In contrast, in the second case,
the vortex dynamics is directly associated with the soliton velocity,
since they move together, except around the collision of solitons
where a more complicated vortex dynamics has been encountered.

In rotating systems, a very accurate multiple-mode (MM) model has been
developed to study the dynamics of condensates within lattices which
forms a system of weakly linked condensates (WLCs)\cite{rot20}.  In
such a work, it has been shown that depending on the geometry of the
condensates, different velocity profiles are imprinted on the
localized on-site functions (LFs). In particular, for an on-site axial
symmetry around the same direction of the rotation axis, the
corresponding LF exhibits a homogeneous velocity field.  For WLCs in
such conditions, it has been shown that the position of vortices along
the low-density paths between sites can be analytically estimated
\cite{je23}. The expression for such a position only involves the
phase difference between sites, instead of the density gradients that
appear in nonrotating systems.  From such an expression, the number of
nucleated vortices when ramping up linearly the rotation frequency, as
done in the experiment of Ref. \cite{wi10}, has been calculated and
shown to be in excellent agreement with Gross-Pitaevskii (GP)
simulations.  In the last year, a renewed interest in vortex dynamics
in rotating condensates has arisen due to the possible relation to
vortex dynamics in stars \cite{Poli2023,Bland2024}. We believe the position of
vortices in such systems could be determined using an analogous
expression, as a result of the distribution of particles inside
the stars. 

The aim of this work is to show that the vortex dynamics is determined
only by the time-dependent phase differences between sites. Then, one
can generate distinct patterns of vortices and make them describe
predetermined types of orbits by appropriately choosing the initial
populations and phase differences between sites of the rotating
lattice.  By constructing two-mode (TM) model Hamiltonians, we will
demonstrate the diverse types of vortex dynamics that can be
engineered for distinct values of the rotation frequency. The time
evolution of the vortices shares the same period as those of Josephson
or self-trapping oscillations in the TM model, thereby facilitating
the tracking of the vortex trajectories.

The paper is organized as follows. In Sec. \ref{sec:Theo} we describe
the theoretical background, which includes a summary of the method for
obtaining the LFs, the description of the trapping potential, a brief
analysis of the phase expressions for the on-site LFs obtained in
Ref. \cite{rot20}. We also analyze the formula for obtaining the
vortex positions when the phase differences and populations evolve in
time \cite{je23}.  In Sec.  \ref{sec:TM}, we construct two
Hamiltonians to obtain the different types of vortex patterns.  We
first describe some characteristics of the hopping parameter entering
the Hamiltonians that determines the equations of motion, and then
analyze the corresponding phase-space diagrams.  In Sec.
\ref{sec:dyn}, we show a variety of vortex dynamics which includes
closed and open orbits for both Hamiltonians. In most cases, estimates
of characteristic times are presented.  In Sec. \ref{sec:conclu}, we
present our conclusions and the Appendix is devoted to exhibit the MM
model equations of motion. There, we also derive a relation between
the hopping parameter and energy differences of stationary states
and discuss other useful parameters.

\section{ \label{sec:Theo}Theoretical framework}

\subsection{On-site localized functions and multimode model }

We first review the construction of the MM model, which is then used
to describe the vortex dynamics in terms of the populations and phase
differences between the sites. An essential requirement for such a
purpose consists in extracting accurate LFs.  In previous works, it
has been described the method for obtaining these localized states
$w_k(\mathbf{r})$ in either nonrotating \cite{mauro4p,cat11} or
rotating \cite{rot20,je23} ring-shaped lattices with $N_s$ sites.
Such LFs are given in terms of stationary states
$\psi _n(\mathbf{r},\Omega)$, which in turn are obtained by solving
the Gross-Pitaevskii (GP) equation,
\begin{equation}
\left[ \hat{H}_0 +
g \, N|\psi_n(\mathbf{r}, \Omega)|^2 - { \Omega}\cdot {\hat{ L}}  \right] \psi_n(\mathbf{r},\Omega)=\mu _n \psi _n (\mathbf{r},\Omega),
\label{GProtstatic}
\end{equation}  
where $n$ is restricted to $-[(N_s-1)/2]\leq n \leq [N_s/2]$
\cite{rot20,je23}.  The operator $\hat{H}_0$ reads
$\hat{H}_0=-\frac{ \hbar^2}{2m}\nabla^2 + V_{\text{t}}({\bf r} )$,
where $V_{\text{t}}({\bf r} )$ is the trapping potential, $g$ is the
3D Rubidium coupling constant, and $\mathbf{\Omega}=\Omega\hat{z}$ is
the applied rotation around the $z$-axis.

From such $N_s$ orthonormal stationary states, the on-site LFs are
obtained through the basis transformation \cite{rot20,je23},
\begin{equation}
w_k({\mathbf r},\Omega)=\frac{1}{\sqrt{N_s}} \sum_{n} \psi_n({\mathbf r},\Omega)
 \, e^{-i n\theta_k } \,,
\label{wannier} 
\end{equation}
where the index $k$ labels the corresponding site, with
$-[(N_s-1)/2]\leq k \leq [N_s/2]$ and $\theta _k=2\pi k/N_s$.

Then, the MM model order parameter can be written using the
orthonormal LFs as,
\begin{equation}
\psi_{\text{MM}} ({\mathbf r},t) = \sum_{k} \,  b_k(t)  \,  w_k ({\mathbf r}, \Omega)
  \,,
\label{orderparameter}
\end{equation}
where $b_k(t)=\sqrt{n_k(t)}e^{i\phi _k(t)}$, and $ N_k = n_k N $ is
the occupation number.  We note that the global phase $\phi _k(t)$
does not represent the total phase in the $k$-site, but it takes into
account only its time dependence, while the spatial profile of the
phase, produced by the rotation, is carried out by the complex LFs,
$w_k(\mathbf{r}, \Omega)$.
   
For obtaining the dynamics, we will use the time-dependent GP equation
in the rotating frame, which reads
\begin{equation}
 \left[ \hat{H}_0 +
g \, N|\psi(\mathbf{r},t)|^2 - \Omega \,\hat{L}_z  \right] \psi(\mathbf{r},t)= i\hbar  \, \frac{\partial\psi(\mathbf{r},t)}{\partial \,t} .
\label{gp}
\end{equation}
Equation (\ref{gp}) will be used to calculate the exact order
parameter $\Psi_{\text{GP}} ({\mathbf r},t) $ and also serves for
deriving the MM model equations of motion and its parameters
\cite{rot20}. For completeness, we include such a MM model approach in
the Appendix.

We finally note that, by inverting the basis transformation in
Eq. (\ref{wannier}), the stationary states in terms of the LFs acquire
the form,
\begin{equation}
\psi_n( \mathbf{r} )=\frac{1}{\sqrt{N_s}} \sum_{k} w_k( \mathbf{r})
 \, e^{i n k 2\pi/N_s} \,.
\label{statn}
\end{equation}
As we are interested in a four-site system, the subscript verifies
$ n \in \{-1;0;1;2\} $.  The stationary states have energies $E_n$
that change their relative values when increasing the rotation
frequency. In addition, the difference between energies with even
(odd) $n$ labels defines the real (imaginary) part of a hopping
parameter (see Appendix).  Each Hamiltonian will involve only the real
or imaginary part of such parameter, and hence their signs will
define the stationary points of the phase-space diagram.  For both
Hamiltonians, such stationary points will correspond to different
stationary states given by Eq. (\ref{statn}).

\subsection{ The trapping potential  }

\begin{figure}[!h]
 \includegraphics[width= 0.9 \columnwidth]{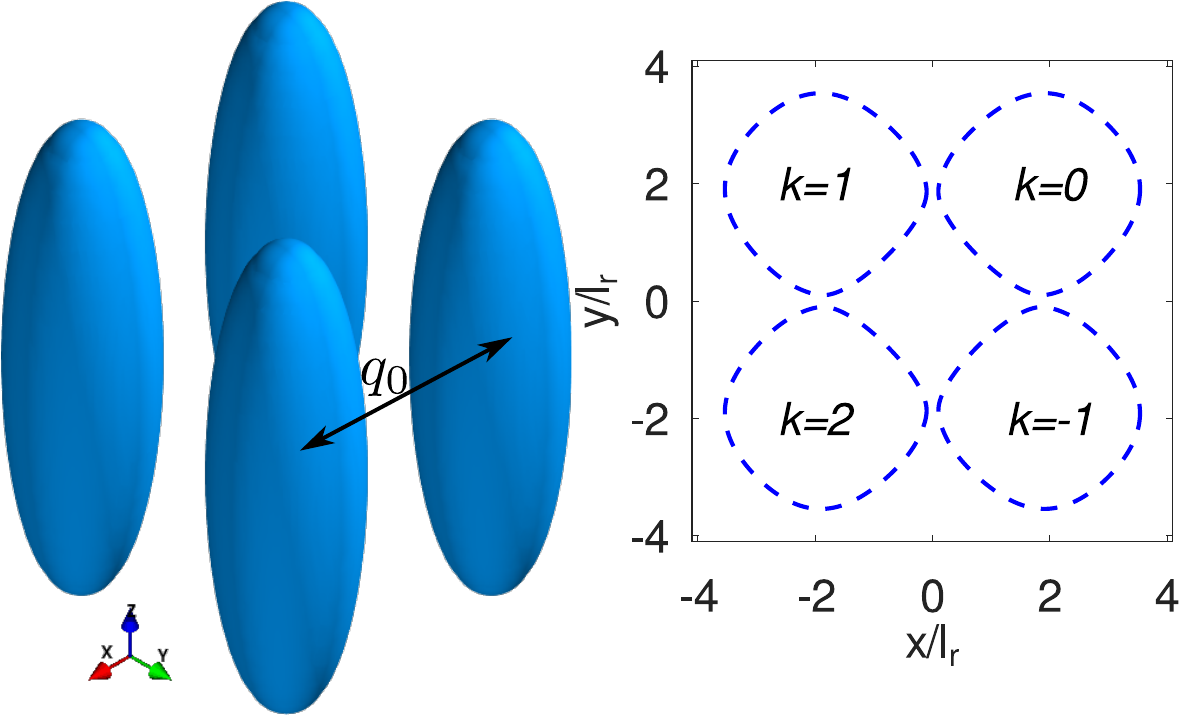}%
 \caption{\label{fig0b} A schematic illustration of the
   $4$-site-lattice condensate is shown in the left panel, where $q_0$
   is the intersite distance. In the right panel, isodensities
   contours are depicted in blue dashed lines, at the $z=0$ plane,
   which correspond to the ground state for $\Omega=0$.  There, we
   also indicate the $k$ value for each site.  }
\end{figure}

In what follows, we will restrict our study to a four-well ring-shaped
trapping potential given by,
\begin{equation}
V_{\text{t}}({\bf r} ) =  \frac{ 1 }{2 }  m  \left[
\omega_{r}^2  r^2  
+ \omega_{z}^2  z^2 \right] 
+   
V_b \left[  \cos^2(\pi x/q_0)+   
 \cos^2(\pi y/q_0)\right],
\label{eq:trap4}
\end{equation}
where $r^2=x^2+y^2$ and $m$ is the atom mass. The harmonic frequencies
are given by $ \omega_{r}= 2 \pi \times 70 $ Hz and
$ \omega_{z}= 2 \pi \times 90 $ Hz, and the lattice parameter is
$ q_0= 5.1 \mu$m $\simeq 3.939 l_r$, with
$l_r=\sqrt{\hbar/(m \omega_r)} $.  Hereafter, the time, energy, and
length will be given in units of $\omega_r^{-1}$, $\hbar\omega_r$, and
$l_r $, respectively.  The parameters of the trapping potential have
been selected similar to those used in the experiment of Albiez
\textit{et al}. \cite{albiez2005} in order to obtain a double-well
system. With the purpose of forming four WLCs whose dynamics can be
treated by the MM model, we further fix the barrier height and number
of Rubidium atoms to $V_b = 25 \hbar\omega_r $ and $ N=10^4 $,
respectively.  In Fig. \ref{fig0b} we show a scheme of the
corresponding WLCs (left panel) and the location of the $k$-sites in
the $x-y$ plane (right panel).

\subsection{ Phases on the on-site localized  functions }

For the trapping potential given by Eq. (\ref{eq:trap4}), it has been
shown \cite{rot20} that the imprinted velocity field within each
localized function turns out to be homogeneous and it is given by
$\mathbf{\Omega}\times\mathbf{r}_{\text{cm}}^k$, where
$ \mathbf{r}_{\text{cm}}^k$ is the corresponding center of mass.  Such
a uniform velocity field is a consequence of the almost circular
symmetry that acquires the on-site-localized density at the $ x-y$
plane for each $z$ value.  In particular, the LF for the $k$-site can
be written as
\begin{equation}
  w_k(\mathbf{r},\Omega) = |w_k(\mathbf{r},\Omega)| e^{i \frac{m}{\hbar}
(\mathbf{r}-\mathbf{r}_{\text{cm}}^k)\cdot(\mathbf{\Omega}\times\mathbf{r}_{\text{cm}}^k)},
  \label{wlphase1}
\end{equation}
where $\mathbf{r}_{\text{cm}}^k $ is calculated using the localized
density $|w_k(\mathbf{r},\Omega)|^2$.  As we will show, the linear
dependence of the argument of $w_k(\mathbf{r},\Omega)$ on the
coordinates turns out to be helpful for determining the dynamics of
vortices along the low-density paths of the lattice.  We may further
rewrite Eq. (\ref{wlphase1}) in terms of the center-of-mass
coordinates: $\mathbf{r}_{\text{cm}}^k = (x_k , y_k , 0)$ as
\begin{equation}
  w_k(\mathbf{r},\Omega) = |w_k(\mathbf{r},\Omega)| e^{i
    \frac{m}{\hbar}(y x_k - x y_k) \Omega}.
  \label{wlphase2}
\end{equation}
For the trapping potential here considered, the coordinates of the
center of mass of the localized densities verify $| x_k|= |y_k|$ and
their absolute values may be approximated by $ q_0/2$.  In the next
section, we will show that such a center of mass is in fact slightly
shifted with respect to the value given by the lattice parameter, due to
the presence of the harmonic confinement and the effect of the
centrifugal force. Disregarding such effects, the LF at the
$k=0$-site can be approximated by
\begin{equation}
w_0(\mathbf{r},\Omega) = |w_0(\mathbf{r},\Omega)| 
 \, e^{i A(-x + y) } \,,
\label{wannier0}
\end{equation}
where the factor in the phase reads
$A = \frac{m}{ 2 \hbar} \, q_0 \, \Omega $.

\begin{figure}[h!]
\tabcolsep=0pt
\begin{tabular}{cc}
\includegraphics[width=0.5\columnwidth]{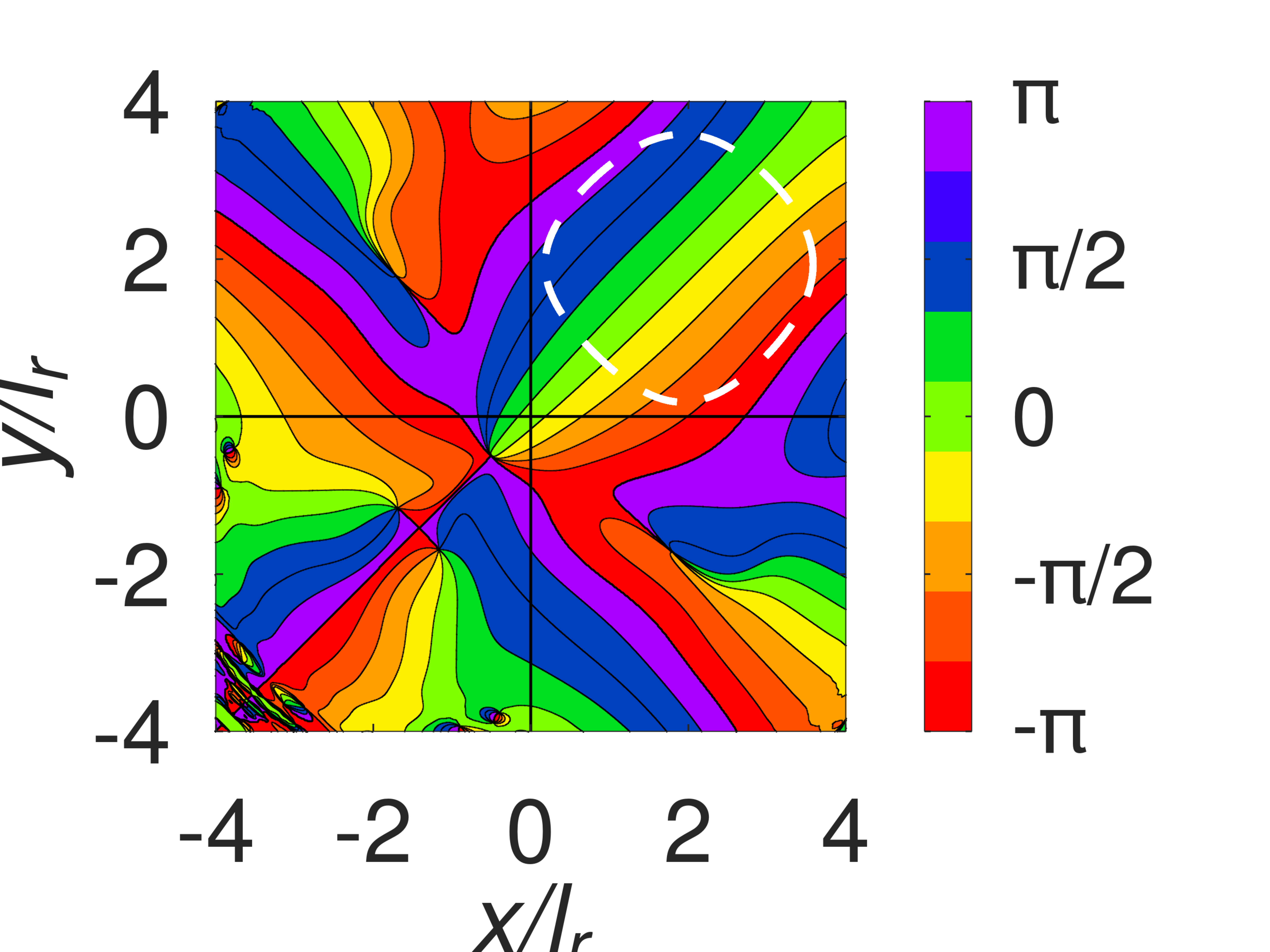}&
\includegraphics[width=0.5\columnwidth]{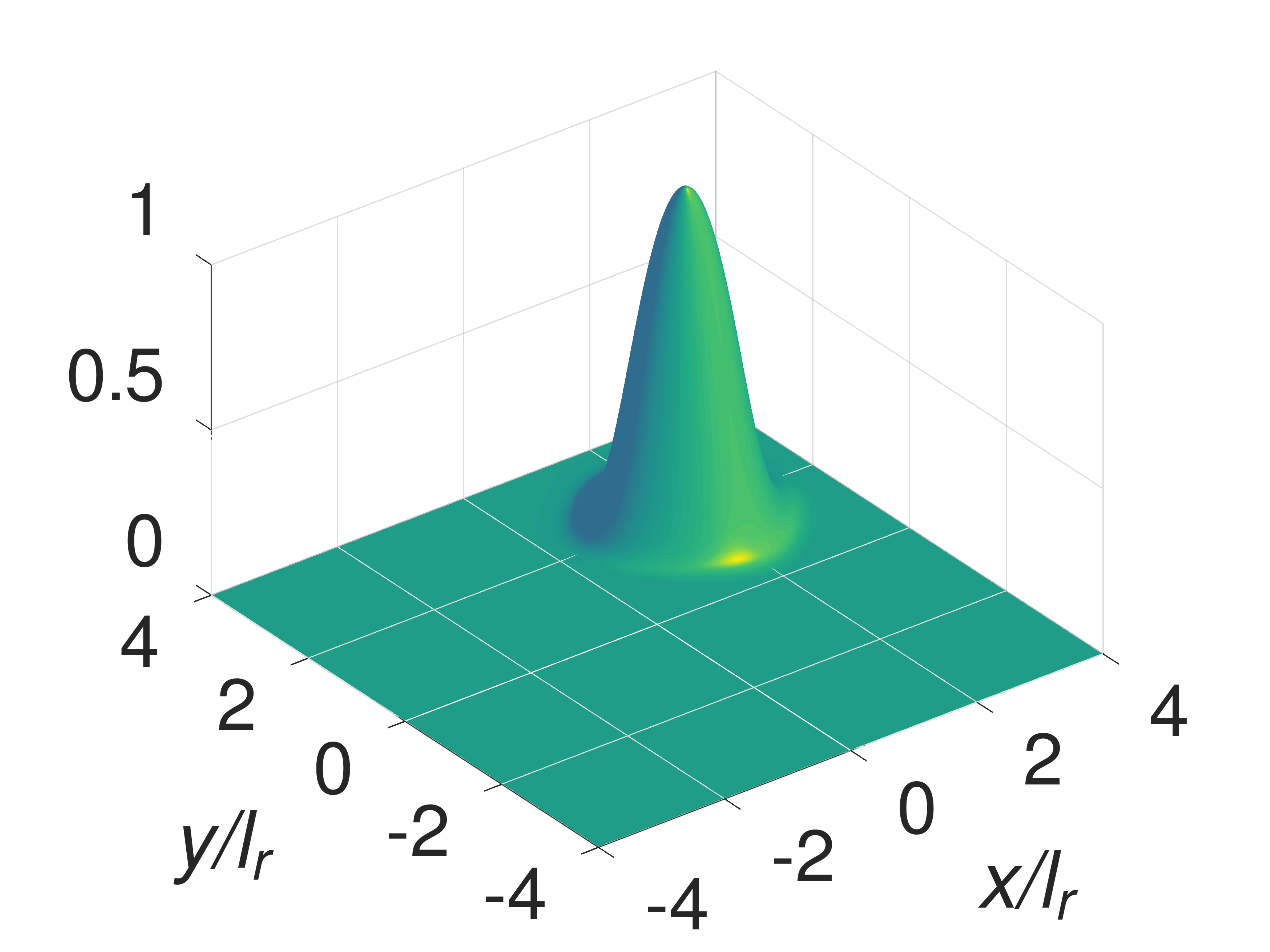}
\end{tabular}                                                            
\caption{\label{figw0} The left panel shows the phase of
  $ w_0(x,y,0) $ for $ \Omega/2 \pi = 45$ Hz.  The white dashed line
  indicates the isodensity contour of such a LF, with a value of
  $10^{-3}$ of the density maximum $ \rho_{\text{max}}$. The color
  scale corresponds to $\arg{( w_0(\mathbf{r})) }$.  In the right
  panel, we depict the renormalized density
  $ |w_0(x,y,0)|^2 / \rho_{\text{max}} $.  }
\end{figure}

In the left panel of Fig. \ref{figw0} we show the phase of
$ w_0(x,y,0) $, where it may be seen it exhibits a linear
coordinate-dependence behavior in accordance to Eq. (\ref{wannier0})
in the corresponding quadrant.  It is important to note that the
approximated expression given by Eq. (\ref{wannier0}) remains valid
around the junctions \cite{je23} connecting neighboring sites.  Then,
as the MM model order parameter is a linear combination of LFs, we
will use the analytic expressions of the phases along the low-density
paths that separate the sites to determine the location of vortices.

\subsection{ Analytical expressions for the vortex coordinates in
  terms of the multimode model variables}

In this section, we briefly outline the method for obtaining the
vortex positions developed for the study of nucleation of vortices in
Ref. \cite{je23}.  We will further show that for a rotating lattice
and according to Eq. (\ref{orderparameter}) the evolution of $ n_k (t)$
and $ \phi_k(t)$ determines the vortex dynamics.  The condition for
the presence of a vortex line with coordinates $(X_v(t), Y_v(t), z) $
leads to a zero value in the time-dependent wavefunction, i.e.,
$ \psi_{\text{MM}} ({ X_v(t), Y_v(t), z, t })=0$. Considering only the
two neighboring LFs with $k=0$ and $k=1$, and approximating their
phases with the expressions of Eq. (\ref{wlphase2}), one obtains the
condition for the vortex coordinates along the low-density path
$x \simeq 0$ for $y>0$
\begin{equation}
  \sqrt{n_0} e^{i \phi_0}  |w_0|  \, e^{i A(-X_v + Y_v) } +   
 \sqrt{n_1} e^{i \phi_1}  |w_1|  \, e^{-i A(X_v+ Y_v) }  = 0   \,.
\label{m01a}
\end{equation}
Hence, defining   $\varphi_1= \phi_1 -\phi_0$,  it leads to
\begin{equation}
  e^{i (  2 A Y_v - \varphi_1 ) }   \sqrt{n_0}   |w_0|   +    \sqrt{n_1}   |w_1|    = 0   \,.
\label{aux}
\end{equation}
Therefore, from the imaginary part of the above equation we find the
following condition
\begin{equation}
    \sin( 2 A Y_v(t)  - \varphi_1 (t) )   = 0   \,,
\label{im}
\end{equation}
which implies that $ 2 A Y_v(t) - \varphi_1(t) =k' \pi$, whereas the
real part of Eq. (\ref{aux}) restricts $k'=2l_y+1$.  Hence, the vortex coordinate
$Y_v(t)$ evolves ruled by the time-dependent phase difference between
the involved sites, following the law
\begin{equation}
Y_v^{(1)} (t)= \left( \frac{\varphi_1(t)}{\pi} + \, 2l_y+1 \right) \, \frac{ \pi
  \hbar}{ m q_0 \Omega} \,,
\label{vort}
\end{equation}
where $l_y $ is an integer number.  Hereafter, the superscript $(k)$
on the coordinates will denote the low-density path between the sites
labeled by $k-1$ and $k$.  We further note that, we will consider
$ -\pi \leq \varphi_1(t) \leq \pi $, which implies $l_y \ge
0$. Then, every time $ \varphi_1(t)$ crosses $\pi $ the value of $l_y $
should be increased by one unit. Such transitions occur at values of $Y_v$
multiples of $ 2 \pi \hbar /( m q_0 \Omega) $. Hence, in the present
study and for rotation frequencies larger than
$\Omega/2 \pi \simeq 28 $ Hz, two values $l_y= 0,1$ are needed for
describing the whole vortex dynamics.

Additionally, from the real part of Eq. (\ref{aux}) one can infer that
$X_v^{(1)}(t)$ should oscillate following the $n_k(t)$
evolution. Indeed, Eq.(\ref{aux}) leads to
$ \frac{ \sqrt{n_0(t)} |w_0({ X_v,Y_v, Z_v})|}{ \sqrt{n_1(t)} |w_1({
    X_v,Y_v, Z_v})| } =1 $, which gives rise to a very small
$X_v^{(1)}$ oscillation amplitude around zero.

We may straightforwardly generalize the expression of the vortex
coordinate given by Eq. (\ref{vort}) to the rest of the junctions by
taking into account that the involved variable $\varphi_k (t)$ may, in
general, take a different value and thus the vortices in other
junctions exhibit distinct dynamics. We may further note that for low
$\Omega $ values, the expression for the vortex coordinates can throw
values $ Y_v \gg q_0 $ and hence it can happen that at some times one
does not observe any vortex inside the lattice.  However, it does not
exist a critical nucleation frequency, given that depending on the
values of $\varphi_k $ a vortex can appear when the phase difference
acquires a value near $\pm \pi$ .

\section{ \label{sec:TM} Symmetric initial conditions: reduction to two  mode model systems  }

In this section, we analyze the phase-space diagrams corresponding to
TM model Hamiltonians obtained by considering that the initial
populations and phase differences alternate their values when moving
around the ring in the counterclockwise direction. In particular, we
restrict to $n_k(0)=n_{k+2}(0) $ and $\varphi_k(0)=\varphi_{k+2}(0)$.
By introducing these conditions into the MM equations of motion given
by Eqs. (\ref{ncmode1hn}) and (\ref{ncmode2hn}), only a single hopping
parameter $K$ and an effective interaction parameter $U_{\text{eff}}$
are involved (see Appendix).  Furthermore, whether
$\varphi_k(0)+\varphi_{k+1}(0)=0$ or
$\varphi_k(0)+\varphi_{k+1}(0)= \pi$, the real part or the imaginary
part of $K$ is retained in such equations, respectively.  By defining
the conjugated coordinates $Z=2(n_0-n_1)=2\Delta N/N$ and
$\varphi=\varphi_1$, one can easily construct both TM model
Hamiltonians.  We will first discuss the dependence of such TM model
parameters on the rotation frequency.

\subsection{  Parameters involved in the model  }

We note that in the static case, the effective on-site interaction
energy is given by a positive real number,
$U_{\mathrm{\text{eff}}}=2.269 \times10^{-3}\hbar\omega_r$, which
remains almost constant for increasing rotation frequencies (see
\cite{mauro4p}).  On the other hand, for nonzero $\Omega$ values the
hopping parameter $K$ becomes a complex number with a decreasing
absolute value as a function of $\Omega$.  For this trapping
potential, it has been shown that the LFs acquire a linear phase which
gives rise to an almost homogeneous velocity field on each site when
the system is subject to rotation \cite{rot20}.  Such a particular phase
profile determines the phase $\Theta$ of the hopping parameter $K$
(see Appendix).  In Fig. \ref{fig0}, the real and imaginary part of
$K$ are depicted as function of $\Omega$, which will define the
distinct types of vortex dynamics of the present study.

\begin{figure}[!h]
 \includegraphics[width=1\columnwidth]{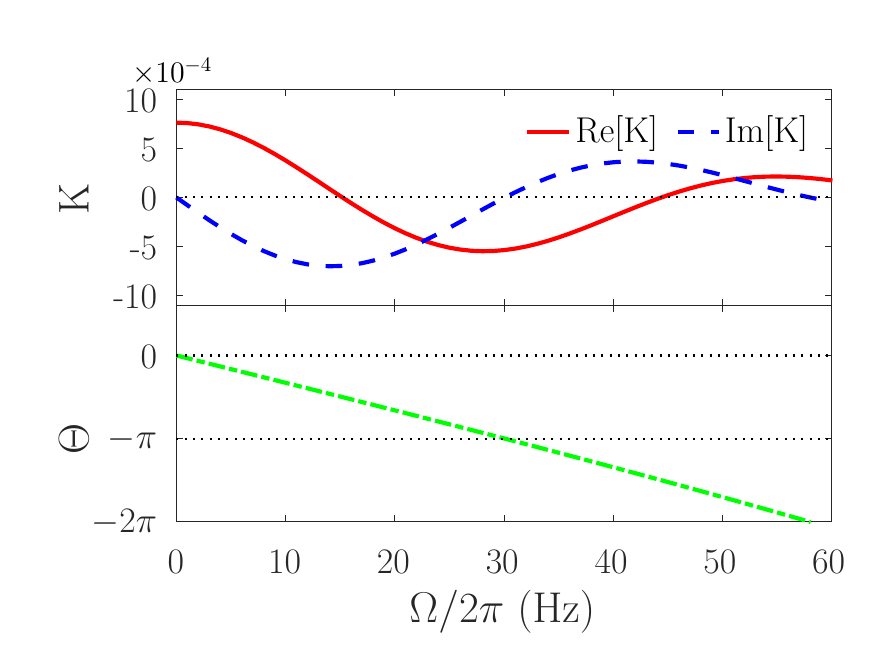}%
 \caption{\label{fig0} Top panel: Real part (red solid line) and
   imaginary part (blue dashed line) of the hopping parameter $K$ as
   function of the rotation frequency. Bottom panel: the argument of
   $K$ as a function of the frequency is depicted as a green
   dashed-dotted line.  }
\end{figure}

Taking into account that $K= |K| e^{i \Theta} $, with
$\Theta = - m \Omega r^2_{\text{cm}} \sin{(2 \pi/N_s)} / \hbar $ (see
Appendix), for our four-site lattice one can roughly approximate the
absolute values of the center-of-mass coordinates in terms of the
intersite distance $|x_k|=|y_k| \simeq q_0/2$, yielding
\begin{equation}
\Theta  \simeq  - \Omega q_0^2 l_r^{-2}/ (2 \omega_r)  \,  .
\label{thetaq0}
\end{equation}
However, in the previous formula we have disregarded the effect of the
harmonic trap, which lowers $ r^2_{\text{cm}}$, and of the centrifugal
force, which slightly increases the $r^2_{\text{cm}}$ of the on-site
localized density with $ \Omega$.  Therefore, a better estimate of
$ r^2_{\text{cm}}$ can be obtained by considering that the net force
applied to the on-site center-of-mass equals zero. Approximating the
lattice potential around its minima to second order in the
coordinates, and equating the sum of the forces given by the harmonic
trap, the lattice potential, and the centrifugal one, it yields
\begin{equation}
| x_{k}|= | y_{k}| \simeq  \frac{ V_b \pi^2 q_0 }{q_0^2  m (\omega_r^2- \Omega^2 ) + 2  V_b \pi^2  } \,  .
\label{xcm}
\end{equation}

\begin{figure}[!h]
 \includegraphics[width=1\columnwidth]{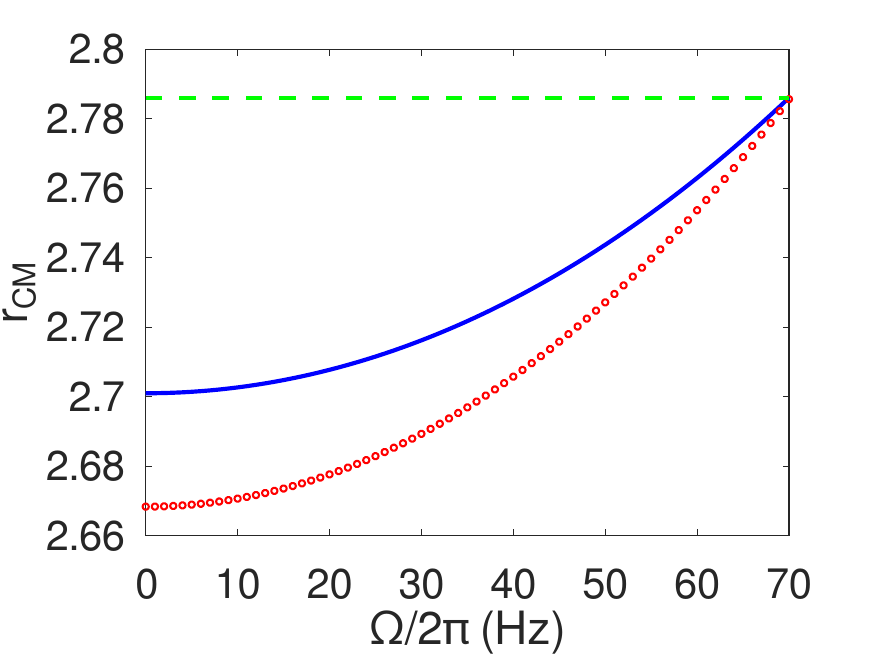}%
 \caption{\label{figeq17} Absolute value of the center-of-mass
   position $r_{\text{cm}}$ as a function of the rotation frequency.
   The exact value obtained using the LFs is depicted with red
   hollow circles.  The solid blue line is the estimate obtained from
   the approximation of Eq. (\ref{xcm}). The rough value, given only
   in terms of the lattice-intersite distance, yields
   $r_{\text{cm}} \simeq q_0/\sqrt{2} = 3.939 \, l_r / \sqrt{2}=
   2.786 \, l_r $ and it is shown as the horizontal green dashed
   line.  }
\end{figure}

In Fig. \ref{figeq17}, it may be seen that the center-of-mass radius
$r_{\text{cm}}$ evaluated using LFs is, indeed, an increasing function
of $\Omega$. Furthermore, when such a rotation frequency reaches the
harmonic trap frequency $\omega_r$, all estimates verify
$r_{\text{cm}} = q_0/ \sqrt{2}$, given that the repulsion of the
centrifugal force cancels the attraction of the harmonic trap.
Moreover, we note that the approximate value obtained through
Eq. (\ref{xcm}) reproduces the exact result within a relative error
less than $0.01$.

\begin{figure}[!h]
 \includegraphics[width=1.1 \columnwidth]{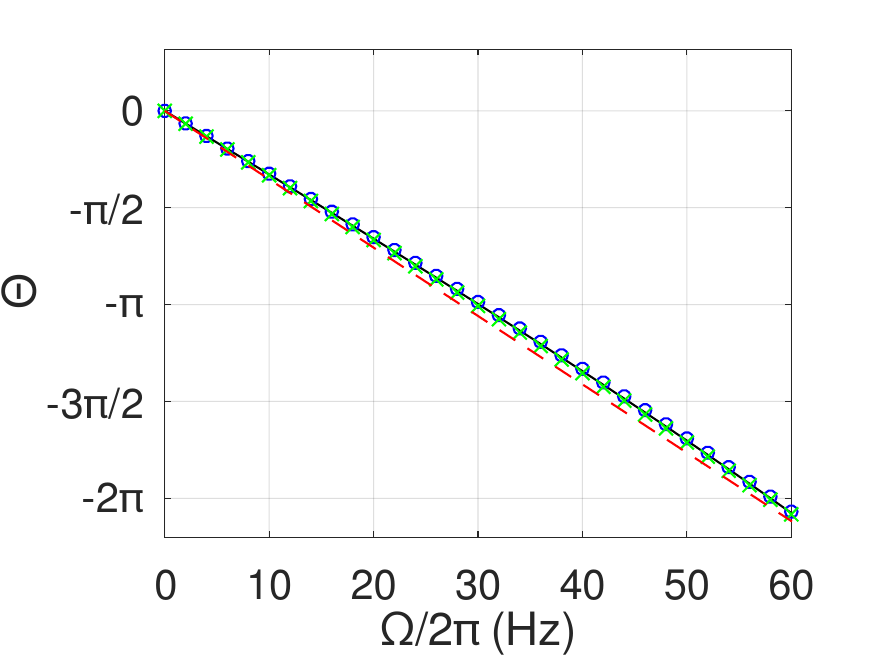}%
\caption{\label{thetacomp} Comparison between different estimates for
  the argument $\Theta$ of the hopping parameter as a function of the
  rotation frequency. The solid black line represents the exact value
  of the argument of $K$. The better estimates are obtained from Eq.
  (\ref{teta}) with the $r_{\text{cm}}$ evaluated by either GP
  simulations (blue hollow circles) or Eq. (\ref{xcm}) (green
  crosses), whereas the less accurate value is given by
  Eq. (\ref{thetaq0}) using $ q_0 = 3.939 \, l_r$ which is shown as a red
  dashed line.  }
\end{figure}

The frequencies where either $\Re(K)$ or $\Im(K)$ vanish are crucial,
since at such frequencies the phase-space diagrams undergo qualitative
changes, as will be discussed in detail in the next section.  In Fig.
\ref{thetacomp}, it may be seen that $\Theta$ as a function of
$\Omega$ slightly deviates from the linear approximation depicted as a
red dashed line.  In particular, using the linear approximation given
by Eq. (\ref{thetaq0}) for searching $\Im(K)=0$, one obtains
$\Omega/2\pi \simeq 28.3$ Hz for $\Theta = -\pi $, whereas the exact
expression yields $\Omega/2\pi \simeq 30$ Hz.  Then, if one wants to
analytically obtain accurate critical rotation frequencies for the
vanishing of $\Re(K)$ and $\Im(K)$, one should take into account the
effect of the forces acting on the center of mass of each site.  From
such a better estimate, one can also obtain the vanishing $\Re(K)$
values, which for $\Theta = -\pi /2$ and $\Theta = - 3 \pi /2$, are
attained around $ \Omega_0/2\pi \simeq 15 $Hz and
$ \Omega_0/2 \pi \simeq 45$ Hz, respectively.

\subsection{ Two-mode model Hamiltonians and phase-space diagrams}

In this section, in order to analyze the phase-space diagram we first
construct two Hamiltonians $ H_i(Z,\varphi) $ with $i= 0, 1 $, making
use of the equations of motion (\ref{ncmode1hn}) and (\ref{ncmode2hn})
restricted to particular initial conditions.  Then, we define
conjugate coordinates $Z$ and $\varphi$ such that the equations of
motion reads,
\begin{equation}
\frac{d\varphi}{dt}=  \frac{\partial H_i}{\partial Z}
\label{varphip0}
\end{equation}
and
\begin{equation}
\frac{dZ}{dt}= - \frac{\partial H_i}{\partial \varphi} \,.
\label{Zp0}
\end{equation}
Given that the critical points, minima, maxima, and saddles of the
Hamiltonians organize the types of regimes inside the phase-space
diagram, we analyze their positions when changing the rotation
frequency.

\subsubsection{ Two-mode model Hamiltonian that involves the real part
  of the hopping}

We first consider the case $\varphi_k(0)+\varphi_{k+1}(0)=0$ in
Eqs.~(\ref{ncmode1hn}) and (\ref{ncmode2hn}).  Given that such
symmetry holds during all the evolution, it has been shown
\cite{nig22} that the equations of motion (\ref{ncmode1hn}) and
(\ref{ncmode2hn}) reduce to that of a two-mode system with Hamiltonian
$H_0$ given by
\begin{equation}
H_0(Z,\varphi) /\gamma_0 = \frac{\Lambda_0}{2} Z^2  - \mathrm{sign}(\Re(K)) \sqrt{1-Z^2} \cos(\varphi) \,,
\label{hamiltoniano}
\end{equation}
for $\Re(K) \ne 0 $, $\Lambda_0= N U_{\text{eff}} / (4 |\Re(K)| )$,
$\gamma_0= { 2 |\Re(K)| }/{\hbar}$. And
\begin{equation}
H_0^{(c)}  = \frac{ N U_{\text{eff}}  }{4 \hbar} \,  Z^2   \,,
\label{hamiltonianoc}
\end{equation}
for $\Re(K) = 0 $, which occurs at a critical  value  $ \Omega =\Omega_0 $.

\begin{figure}[h!]
  \centering
  \includegraphics[width=0.7\columnwidth,clip=true]{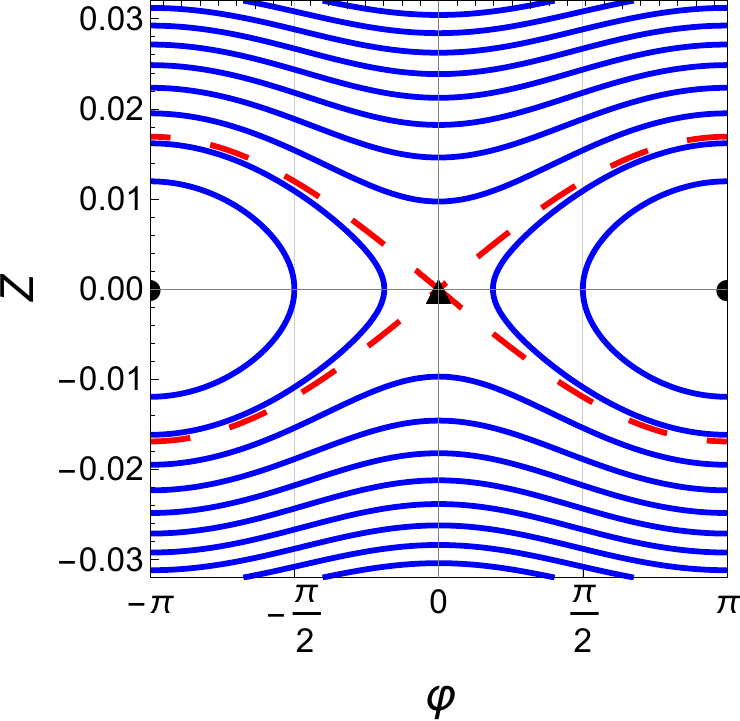}\\
  \includegraphics[width=0.7\columnwidth,clip=true]{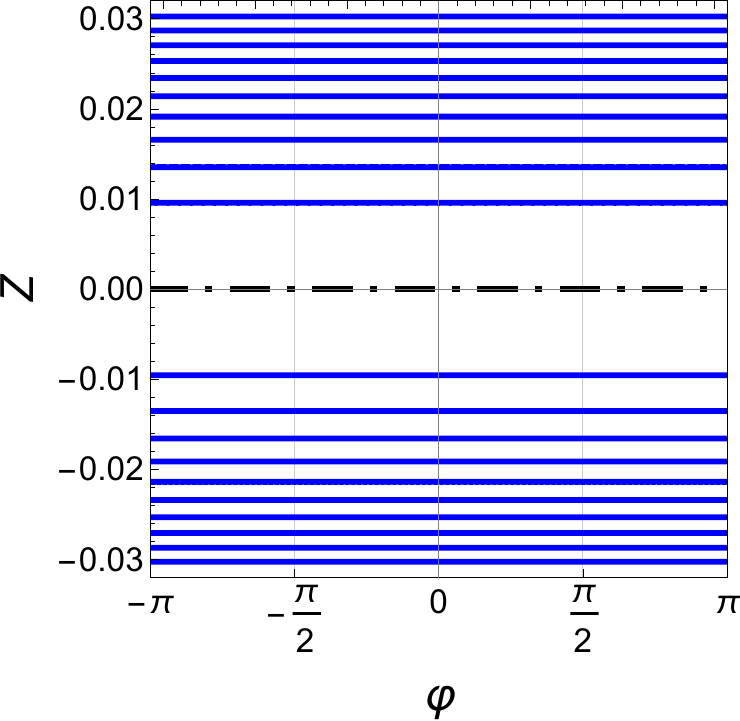}\\
  \includegraphics[width=0.7\columnwidth,clip=true]{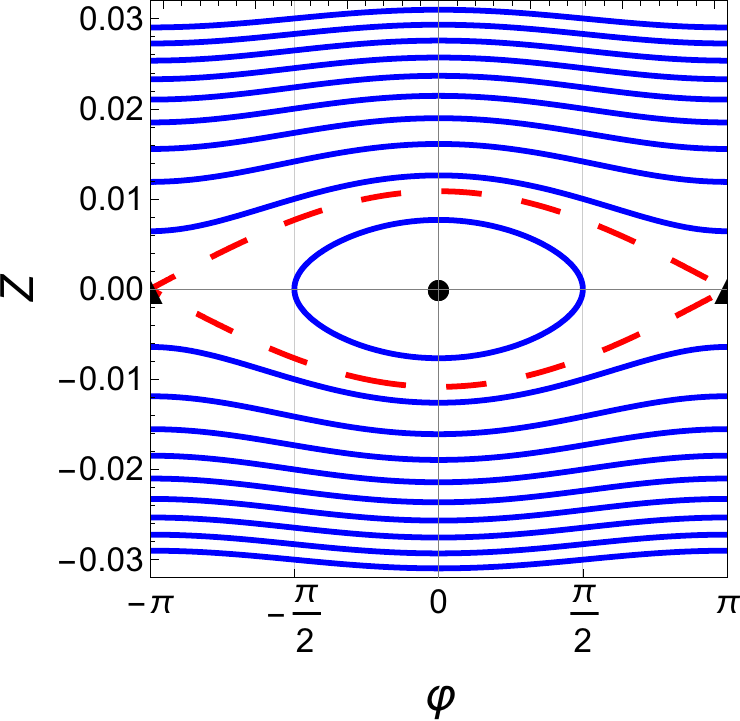}
  \caption{\label{diagrama35} Phase-space diagrams of $H_0$ for
    $ \Omega/ 2 \pi = 35 $ Hz (top panel), for the critical frequency
    $ \Omega_c/2 \pi = 45 $ Hz (middle panel), and for
    $ \Omega/ 2 \pi = 50 $ Hz (bottom panel).  The TM orbits are
    depicted with solid blue lines and the separatrices between closed
    and open orbits with dashed red lines.  The stationary points are
    indicated with a dot (minimum) and with a filled triangle which
    corresponds to a saddle point.  The black dot-dashed line in the
    middle panel indicates a one dimensional variety of stationary
    points.  }
\end{figure}

In the latter case, the critical values in the range of the rotation
frequencies studied (see Fig. \ref{fig0}) are
$\Omega^{(1)}_0/2\pi \simeq 15$ Hz and
$ \Omega^{(2)}_0/2 \pi \simeq 45$ Hz.  For
$0 \le \Omega< \Omega^{(1)}_0$ and $ \Omega^{(2)}_0< \Omega $ the
phase-space diagrams exhibit a Hamiltonian minimum at $Z=0$ and
$\varphi=0$, around which, closed orbits are present. Such orbits are
called plasma- or $0$-modes. A saddle point located at $Z=0$ and
$\varphi=\pi$ defines the Hamiltonian value for the separatrix between
the closed and open orbits. The open orbits exhibit a running phase
difference typical of the self-trapping (ST) regime, which is
characterized by an imbalance $Z$ that never vanishes.  For the lower
frequency interval, the corresponding phase-space diagram has been
studied previously in \cite{nig22}. Here, we show the phase-space
diagram for $\Omega/2 \pi =50$ Hz in the bottom panel of
Fig. \ref{diagrama35}.  In the rotation frequency interval
$ \Omega^{(1)}_0 < \Omega< \Omega^{(2)}_0$, the Hamiltonian minimum
moves to $Z=0$ and $\varphi=\pi$ around which closed orbits called
$ \pi-$modes are present, whereas $Z=0$ and $\varphi=0$ becomes the
new saddle point (top panel of Fig. \ref{diagrama35}).  For both
plasma and $\pi$ modes, the maximum difference of the particle numbers
is $\Delta N_c = N \sqrt{\Lambda_0 -1}/\Lambda_0 $, which is attained
at $\varphi=0$ and $\varphi=\pi $, respectively.

The MM states associated to the dynamics  restricted to
the  symmetry that gives rise to $H_0$ do not exhibit vortices at the
center of the $x-y$ plane, except for $Z=0$ and $\phi= \pi$.  This can
be verified using the expression of the order parameter with the four
localized functions, which at $x=y=0$ yields
\begin{equation}
\psi_{\text{MM}}= 2 |w_0({ 0,0, z})| \left(\sqrt{n_0} e^{i\phi_0} + \sqrt{n_1}
e^{i\phi_1} \right) \, ,
\label{enori}
\end{equation}
where we have used that $ |w_k ({ 0,0, z})| = |w_0({ 0,0, z})| $.
From the previous expression, one can assure that
\begin{equation}
\psi_{\text{MM}}= 2 |w_0({ 0,0, z})|\, e^{i\phi_0} \left(\sqrt{n_0} + \sqrt{n_1}
e^{i\varphi} \right) \ne 0 \, ,
\label{enori1}
\end{equation}
except for $\varphi=\pi$ and $Z=0$, which is always a stationary point
and hence it does not participate in any dynamics.

It is interesting to note that the stationary points of the
phase-space diagrams at $\varphi=0$ (bottom panel of
Fig. \ref{diagrama35}) and $\varphi=\pi $ (top panel) correspond to
the GP-stationary states $\psi_0( \mathbf{r} )$ and
$ \psi_2( \mathbf{r} )$, respectively, as it can be seen from
Eq. (\ref{statn}). Hence, the minimum is related to different $n$
values depending on the selected range of the rotation
frequency. Although the central, doubly quantized vortex at the
$z$-axis does not survive in the surrounding region of the phase-space
diagram, we will see that it is connected to a two-vortex dynamics
around such an axis.

\subsubsection{ Two-mode model Hamiltonian that involves the imaginary part of the hopping}

A different type of dynamics arises if we consider $\varphi_k(0) + \varphi_{k+1}(0) =
\pi$.  In such a case, the equations of motion (\ref{ncmode1hn}) and (\ref{ncmode2hn})
give rise to the following TM model Hamiltonian,
\begin{equation}
H_1 /\gamma_1 = \frac{\Lambda_1}{2} Z^2 + \mathrm{sign}(\Im(K)) \sqrt{1-Z^2}
\sin(\varphi) \,,
\label{hamiltonianov}
\end{equation}
for $\Im(K) \ne 0 $,  $\Lambda_1= N U_{\text{eff}} / (4  |\Im(K)| )$,  and
 $\gamma_1= { 2 |\Im(K)| }/{\hbar}$.

 In the particular case of $\Im(K) = 0 $, one has the same expression
 for the critical Hamiltonian as that given by
 Eq. (\ref{hamiltonianoc}),
\begin{equation}
H_1^{(c)}  = \frac{ N U_{\text{eff}}  }{4 \hbar} \,  Z^2   \,.
\label{hamiltonianocv}
\end{equation}
%

\begin{figure}[h!]
  \centering
  \includegraphics[width=0.7\columnwidth,clip=true]{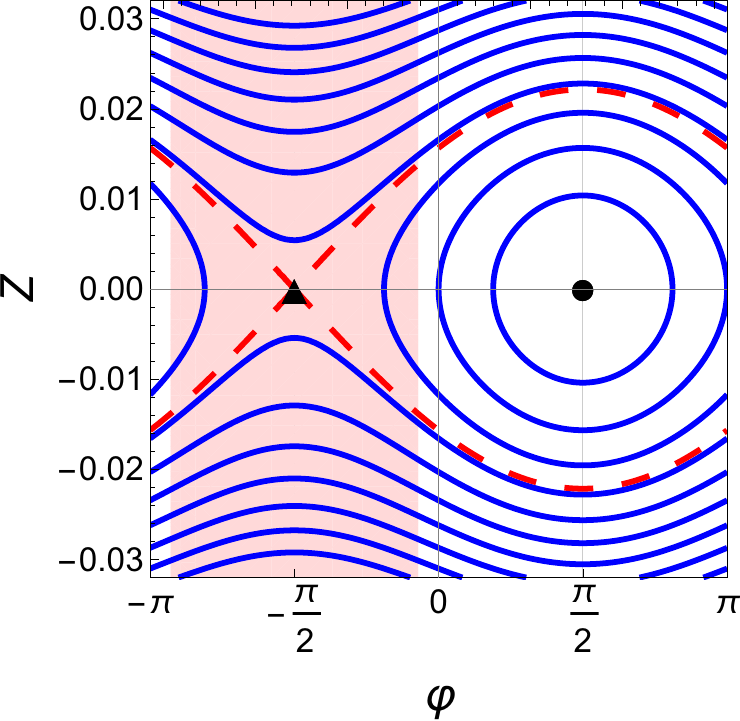}\\
  \includegraphics[width=0.7\columnwidth,clip=true]{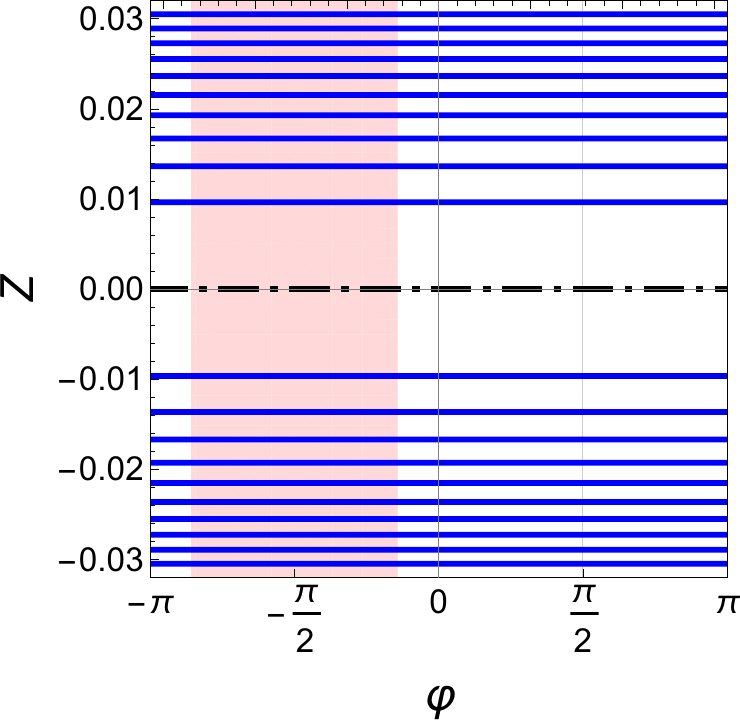}\\
  \includegraphics[width=0.7\columnwidth,clip=true]{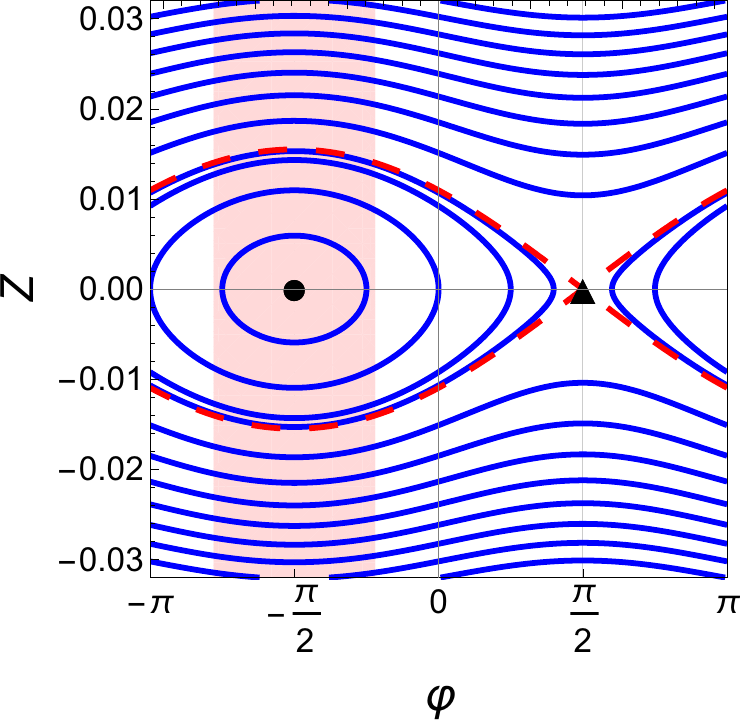}
  \caption{\label{diagrama15} Phase-space diagrams of $H_1$ for
    $ \Omega/ 2 \pi = 15 $ Hz (top panel), for the critical frequency
    $ \Omega_c/2 \pi = 30 $ Hz (middle panel), and for
    $ \Omega/ 2 \pi = 45 $ Hz (bottom panel).  The TM orbits are
    depicted with solid blue lines and the separatrices between closed
    and open orbits with red dashed lines.  The stationary points are
    indicated with a dot (minimum) and with a filled triangle which
    corresponds to a saddle point. The black dot-dashed line in the
    middle panel indicates a one dimensional variety of stationary
    points. The pink-shaded areas indicate the regions with the
    presence of a central antivortex.}
\end{figure}

The condition $\Im(K) = 0 $ is attained around
$\Omega_1 /2\pi \simeq 30$ Hz.  For $0 \le \Omega< \Omega_1$, the
phase-space diagram exhibits a Hamiltonian minimum at $Z=0$ and
$\varphi=\pi/2 $.  Around such a point, the closed orbits , which can
be viewed in the upper panel of Fig. \ref{diagrama15}, will be
referred to as $\pi/2$-modes. The saddle point located at $Z=0$ and
$\varphi=-\pi/2 $ determines the Hamiltonian value that defines the
separatrix (depicted with red dashed lines) between the closed orbits
and the open ones that belong to the ST regime.  In the middle panel
of Fig. \ref{diagrama15}, it may be viewed that $Z$ remains
constant. In particular, the black dot-dashed line at $Z=0$ represents
a set of stationary points.  For rotation frequencies
$ \Omega_1 < \Omega$, the Hamiltonian minimum turns out to be at $Z=0$
and $\varphi=-\pi/2$, see bottom panel of
Fig. \ref{diagrama15}. Around such a point, closed orbits are present,
which will be called $ -\pi/2$-modes. The saddle point in this case is
located at $Z=0$ and $\varphi= \pi/2$.

For both the $\pi/2$-mode and the $-\pi/2$-mode the maximum difference
of the particle numbers is
$\Delta N_c = N \sqrt{\Lambda_1 -1}/\Lambda_1 $, which is attained at
$\varphi=\pi/2$ and $\varphi=-\pi/2 $, respectively. For larger
imbalances, the ST dynamics develops.

The Hamiltonian $H_1$ is related to systems with a permanent central
vortex, and hence the velocity circulation around the lattice is
nonzero for any rotation frequency. However, the central vortex can
invert its charge during the dynamics.  In particular, the phase-space
diagrams become partitioned into regions with either a vortex or an
antivortex.  In the phase-space diagrams of Fig. \ref{diagrama15}, we
show as pink-shaded areas the regions where an antivortex is present
along the $z$-axis. Hence, when crossing the borders of such areas,
creation or annihilation of vortex-antivortex pairs occurs to
produce the change of the central vortex charge.

In order to prove the existence of such a central vortex, we can
examine the MM model order parameter at the $z$ axis using that
$|w_k(0,0,z)|=|w_0(0,0,z)|$. There, it can be written as
\begin{multline}
\psi_{\text{MM}}=    |w_0|   \left(\sqrt{n_0} + \sqrt{n_1}  e^{i\varphi} + \sqrt{n_0}  e^{i\pi}  + \sqrt{n_1}  e^{i(\varphi+\pi)} \right),\\
\label{enoria}
\end{multline}
which vanishes for all $\varphi$ and $Z$ values. Then, we conclude
that a central vortex is always present for any time evolution, while
the actual charge of such a vortex depends on the region of the
phase-space diagram where the orbit passes through.

The stationary points of the phase-space diagrams for this Hamiltonian
correspond to the GP-stationary states $\psi_1( \mathbf{r} )$ and
$ \psi_{-1}( \mathbf{r} )$, for $\varphi=\pi/2$ and $\varphi=-\pi/2 $,
respectively, as it can be shown by comparing $\psi_{\text{MM}}$ of
Eq. (\ref{orderparameter}) with equal populations and the condition on
the phases, with Eq. (\ref{statn}). The state with $n=1$ ( $n=-1$)
corresponds to a central vortex (antivortex) state; hence, its
corresponding coordinates lie within the non shaded (shaded) area of
Fig. \ref{diagrama15}. It is interesting to note that the shaded areas
become narrower when the rotation frequency increases, thus
decreasing the probability of finding such central antivortices.

\section{\label{sec:dyn} Vortex dynamics }

\subsection{ Vortex dynamics at the critical frequencies}

We will analyze the dynamics related to the Hamiltonian $H_0^{(c)}$
and $H_1^{(c)}$ where either $ \Re(K)=0$ or $ \Im(K)=0$, respectively.
In both cases we have
$ \dot{\varphi}= \frac{ N U_{\text{eff}} }{ 2 \hbar} Z $ and
$ \dot{Z}=0$. Then, if $ \varphi(0)=0$,
\begin{equation}
  \varphi(t)= \frac{ N U_{\text{eff}} }{ 2 \hbar} Z t = \frac{ \Delta N U_{\text{eff}} t }{\hbar}.
\label{varphit}
\end{equation}
Hence, the time period $\tau$ for the phase difference to increase
$ 2\pi$ yields
\begin{equation}
\tau = \frac{  2 \pi \hbar}{ \Delta N U_{\text{eff}}  }.
\label{tauphi}
\end{equation}
We note that, as $ \dot{\varphi}(t) \ne 0 $ during all the evolution,
the vortex trajectories do not exhibit turning points
(cf. Eq.~(\ref{vort})).

Another relevant time interval is the time $T_t$ that a vortex spends
in traveling from the trap center to outside the lattice, which can be
estimated by considering the lattice limits as $ |y| < q_0$, and using,
\begin{equation}
Y^{(1)}_v(t)= \left( \frac{ \Delta N U_{\text{eff}} t }{ \hbar \pi}  + \, 2l+1  \right) \, \frac{  \pi \hbar}{ m q_0 \Omega} \,,
\end{equation}
which yields
\begin{equation}
T_t      \simeq   \frac{   m q_0^2  \Omega}{ \Delta N U_{\text{eff}} }  \,.
\label{timet}
\end{equation}
In contrast to $\tau$,  this traveling time increases
linearly with $ \Omega$.

\subsubsection{ Dynamics  related to the  real part of the hopping, with even number of vortices}

Due to the symmetry of the system and the initial conditions, we can
restrict the study of the vortex dynamics to that developed along two
low-density paths. We focus on the paths identified by the coordinates
superscript $ (k)$ with $k =0 $ and $k=1$, which correspond to the
positive semiaxes $x$ and $y$, respectively.  Then, for the
Hamiltonian $H_0^{(c)}$, where $\Re(K)=0$, using vanishing initial
phase differences, such vortex coordinates acquire the form
\begin{equation}
Y^{(1)}_v(t)= \left( \frac{ \Delta N U_{\text{eff}} t }{ \hbar \pi}  + \, 2l_y+1  \right) \, \frac{  \pi \hbar}{ m q_0 \Omega} \,,
\label{vortcy1}
\end{equation}
at the junction (1), and 
\begin{equation}
X^{(0)}_v(t)= \left( \frac{ -\Delta N U_{\text{eff}} t }{ \hbar \pi}  + \, 2l_x+1  \right) \, \frac{  \pi \hbar}{ m q_0 \Omega} \,,
\label{vortcx1}
\end{equation}
at the junction (0). Hence, taking into account the present symmetry
for the dynamics along the other low-density paths, one has
$ X^{(2)}_v(t)= - X^{(0)}_v(t) $ and $ Y^{(-1)}_v(t)= -
Y^{(1)}_v(t)$. Then, if $N_0>N_1$ the vortices enter with constant
velocity along the $x$-axis and depart from the system along the
$y$-axis.

\begin{figure}[!h]
  \includegraphics[width=1\columnwidth]{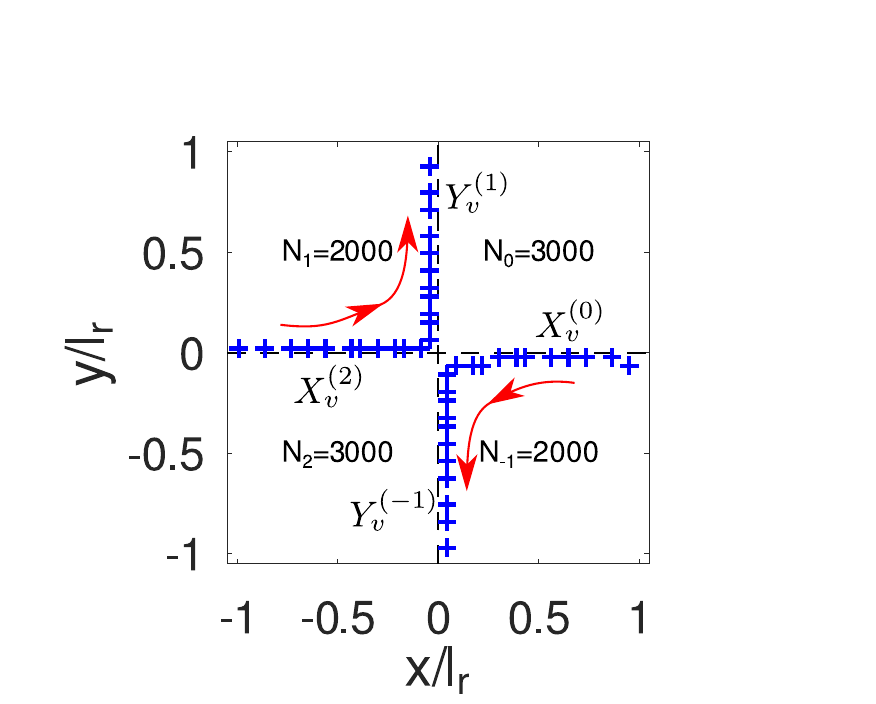}%
  \caption{\label{fig6} Trajectories of two vortices near the trap
    center. The coordinates are extracted from
    $ \psi_{\text{MM}} ({\mathbf r},t)$ using a plaquette method and
    are marked with blue plus signs.  We use a large $\Delta N=1000$
    value and a rotation frequency $ \Omega_c / 2 \pi = 15 $ Hz. The
    arrows indicate the vortex velocity direction in each low-density
    path. Such density paths are indicated by the relevant coordinate
    with its superscript. }
\end{figure}

In Fig. \ref{fig6} we show the vortex trajectories in the $x-y$ plane,
where the positions have been extracted from
$ \psi_{\text{MM}} ({\mathbf r},t)$ of Eq. (\ref{orderparameter})
using the plaquette method of Ref. \cite{foster10}.  It is interesting
to note that, although the incoming vortices approach the $z$-axis at
the same time, they never form a doubly quantized vortex since the
trajectories are well separated at the origin (see Fig. \ref{fig6}).
For such a figure we have chosen a large imbalance $\Delta N= 1000$ to
amplify this behavior.  In particular, it may be seen that the vortex
coming from the positive $x$ axis, $ X^{(0)}_v(t)$, is slightly pushed
to negative $y$ values, and since the vortex then moves away along the
$y<0$ junction, given by $ Y^{(-1)}_v(t)$, it never touches the $z$
axis. On the other hand, the incoming vortex from the negative $x$
values $ X^{(2)}_v(t)$ moves away along the $y>0$ junction, yielding
$ Y^{(1)}_v(t)$.  Such shifts from the axes are in accordance with the
condition
$ \sqrt{n_0(t)} |w_0({ X_v,Y_v, Z_v})| = \sqrt{n_1(t)} |w_1({ X_v,Y_v,
  Z_v})| $ derived from the real part of Eq. (\ref{aux}), given that
$ {n_0(t)} > {n_1(t) } $, it implies that the vortex is nearer the
$k=1$-site than the $k=0$ one.

\begin{figure}[!h]
  \centering
  \includegraphics[width=0.8\columnwidth,clip=true]{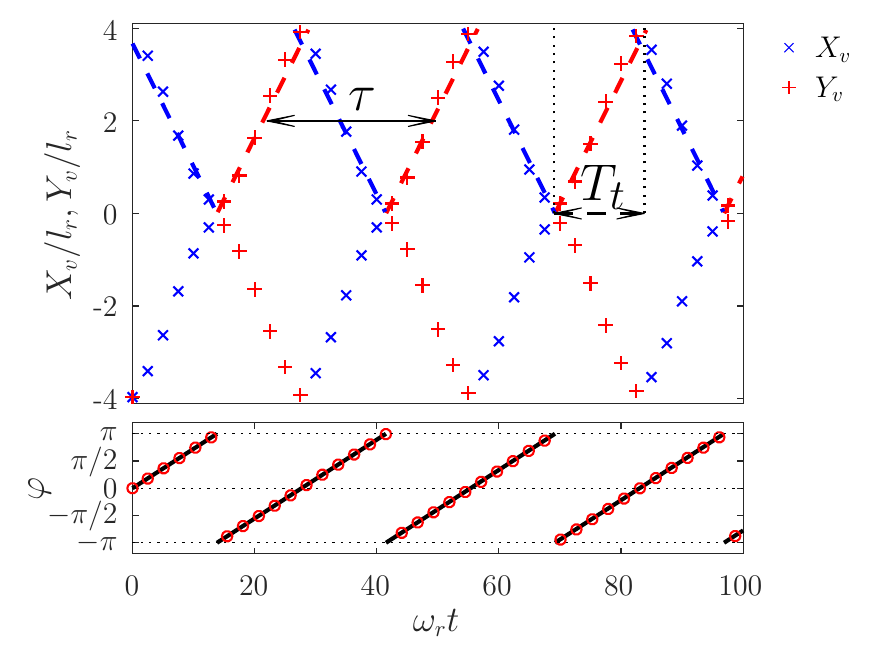}\\
  \includegraphics[width=0.8\columnwidth,clip=true]{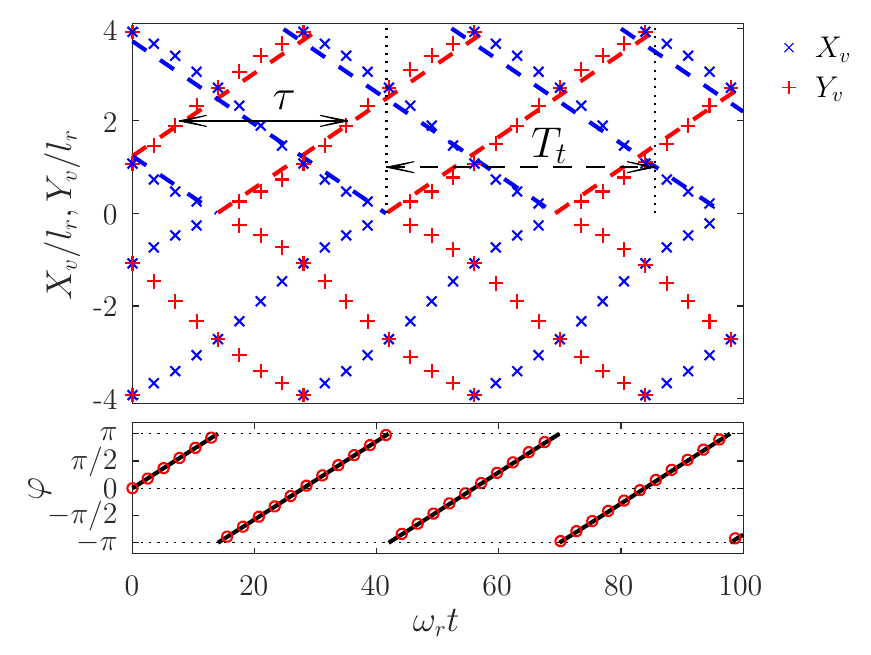}
  \caption{\label{figom15} Vortex coordinates as functions of time
    extracted from $\psi_{\text{MM}} ({\mathbf r},t)$ for
    $ \Omega_c / 2 \pi = 15 $ Hz (top panel) and
    $ \Omega_c / 2 \pi = 45 $ Hz (bottom panel) with $\Delta N=100$
    and $\varphi(0)=0$.  The blue crosses and red plus signs of the
    positive values indicate the coordinates $ X^{(0)}_v(t)$ and
    $Y^{(1)}_v(t)$, respectively.  The same symbols are utilized for
    the negative values which correspond to $ X^{(2)}_v(t)$ and
    $Y^{(-1)}_v(t)$. The dashed red and blue lines represent the
    predicted values from Eqs. (\ref{vortcy1}) and (\ref{vortcx1}),
    respectively.  The horizontal lines indicate the period
    $ \tau = 27.6 \omega_r^{-1} $ (solid) and the traveling times
    (dashed) $T_t = 14.8 \omega_r^{-1}$ (top)
    $T_t = 43.9 \omega_r^{-1}$ (bottom). Below each panel, the phase
    difference $\varphi(t)$ is depicted using the MM model given by
    Eq. (\ref{varphit}) as a solid black line and the GP values are
    shown as red circles. The GP phase difference is calculated from
    $ \Psi_{\text{GP}} ({\mathbf r},t)$ as the difference of the
    phases between the centers of the corresponding sites. }
\end{figure}

Hereafter, in this section we will work with $\Delta N=100$, so that
any shift from the axes becomes negligible. We note that, this
imbalance yields a smaller time period, which allows us to better
follow the dynamics.

The result for the vortex dynamics at two values of the critical
frequency $\Omega_c$ is depicted in Fig. \ref{figom15},
where the time evolution of the vortex positions $ X^{(0)}_v(t)$ ,
$Y^{(1)}_v(t)$, $ X^{(2)}_v(t)$ , and $Y^{(-1)}_v(t)$, along each
semiaxis are shown as function of time for $\Omega_c/2\pi=15$Hz (top
panel), and $\Omega_c/2\pi=45$Hz (bottom panel). The initial condition
$\varphi(0)=0$ implies that the initial coordinate reads
$ Y^{(1)}_v(0) \simeq 3.7 \, l_r$ and
$ Y^{(1)}_v(0) \simeq 1.2 \, l_r$, for the top and bottom panels,
respectively.  The population imbalance is $ \Delta N=100$ which
yields a time period, given by Eq. (\ref{tauphi}),
$ \tau = 27.6 \omega_r^{-1}$. Such a period can be viewed in the
vortex dynamics as the time interval between two successive vortices
pass by the same point of the $ x-y$ plane.  In Fig. \ref{figom15} it
may be seen that the theoretical value is in good accordance with the
evolution in both panels.  The traveling times given by
Eq. (\ref{timet}) for the top and bottom panels yield
$T_t=14.8 \omega_r^{-1}$ and $T_t = 43.9 \omega_r^{-1}$, respectively,
which implies that the vortices inside the lattice move slower when
the rotation frequency is increased.  Below each panel one can observe
a perfect accordance of the phase difference $\varphi(t)$ obtained by
the MM model and its corresponding GP simulation.  Such type of
open orbits are obtained also in the self-trapping regime, where a
running phase is developed.  However, in such a case the vortex
trajectories exhibit a little oscillation in the perpendicular
direction due to the imbalance oscillation.

\subsubsection{Dynamics that involves the imaginary part of the
  hopping, with odd number of vortices}

\begin{figure}[!h]
 \includegraphics[width=1\columnwidth]{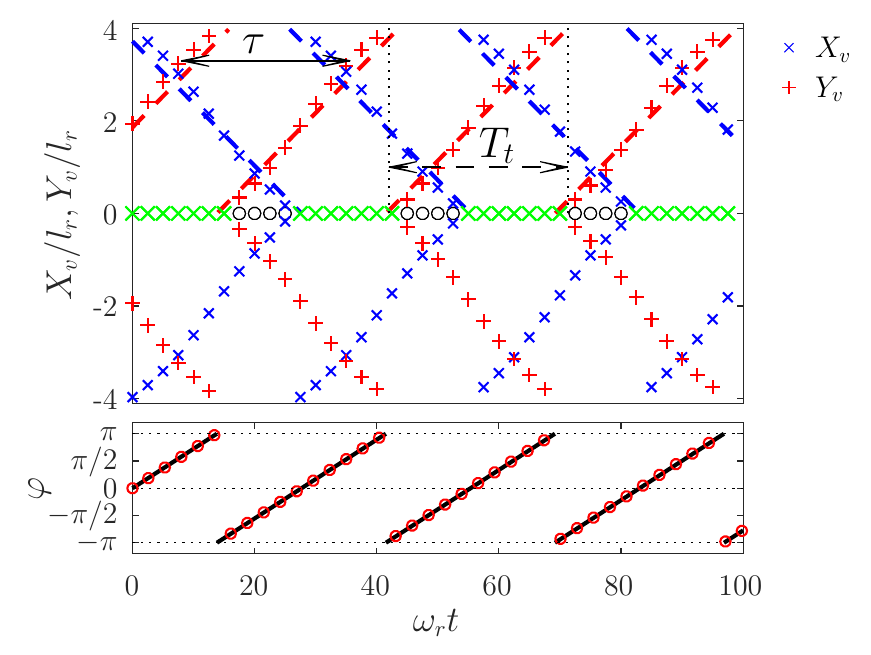}%
 \caption{\label{figom30} 
Vortex  coordinates as  functions of time  extracted from
  the    $\psi_{\text{MM}} ({\mathbf r},t)$ 
  using the plaquette method are shown in the top panel.  The
  blue crosses and red plus signs  of the positive coordinate
  values   indicate $ X^{(0)}_v(t)$ and $Y^{(1)}_v(t)$,
  respectively, whereas  the same symbols are utilized for the
  negative values which correspond to  $ X^{(2)}_v(t)$ and
  $Y^{(-1)}_v(t)$.  The green crosses  and the black circles
  indicate the central vortices  and antivortices, respectively.
  The dashed red  and blue lines represent the predicted values
  from Eqs. (\ref{vortcyv}) and (\ref{vortcxv}),
  respectively.  The
  horizontal lines  indicate the period $ \tau =  27.6
  \omega_r^{-1} $ (solid)  and  the traveling time $T_t= 29.3
  \omega_r^{-1}$ (dashed).
  The initial conditions correspond to $\Delta N=100$ and $\varphi=0$,
  for $ \Omega_c/ 2 \pi = 30 $ Hz.  In the bottom panel the
  corresponding phase difference $\varphi(t)$ is shown as a function
  of time. The results extracted from GP simulations and the MM model
  are depicted with red circles and a solid black line,
  respectively. }
\end{figure}

The dynamics related to the Hamiltonian $H_1^{(c)}$, where
$ \Im(K)=0$, have substantial differences with the previous one.
First, we recall that a central vortex or an antivortex is always
present along the $z$-axis.  If the initial phase difference at the
junction (1) is zero, the junction (0) should exhibit initially a
$\pi$ phase difference, since $\varphi_k + \varphi_{k+1}= \pi$ for
$H_1$. Therefore, the motion of the vortices are described by,
\begin{equation}
Y^{(1)}_v(t)= \left( \frac{ \Delta N U_{\text{eff}} t }{ \hbar \pi}  + \, 2l_y+1  \right) \, \frac{  \pi \hbar}{ m q_0 \Omega} \,,
\label{vortcyv}
\end{equation}
at the low-density path  (1), and 
\begin{equation}
X^{(0)}_v(t)= \left( \frac{ -\Delta N U_{\text{eff}} t }{ \hbar \pi}  + \, 2l_x  \right) \, \frac{  \pi \hbar}{ m q_0 \Omega} \,,
\label{vortcxv}
\end{equation}
at the low-density path (0).  Along the negative semiaxes one has
$ X_v^{(2)} (t)= - X_v^{(0)}(t)$ and
$ Y_v^{(-1)} (t)= - Y_v^{(1)}(t)$.  From Eqs.  (\ref{vortcyv}) and
(\ref{vortcxv}) one can see that the coordinates vanish at different
times, implying that a vortex departs from the trap center along the
$y$-axis before another enters along the $x$-axis. Then, in the whole
lattice, when two vortices leave the trap center along the $y$- axis,
the central vortex becomes an antivortex until the other two vortices
reach the central zone coming from the $x$-axis, as shown in
Fig. \ref{figom30}.  Such a process necessarily involves the creation
and then the annihilation of a vortex-antivortex pair to locally
conserve the velocity-field circulation.

\subsection{Vortex dynamics related to the  closed orbits in the phase-space diagrams}

The stationary points on the phase-space diagram also define
stationary vortex states if the corresponding vortex coordinates are
inside the lattice edge, which is the case for high enough
frequencies. In particular, the minima of the Hamiltonian that are
associated with closed orbits in the phase-space diagram give rise to
closed orbits around the stationary vortex.  For such vortices, the
small-oscillation period is the same as that of the Hamiltonian
orbits.  Therefore, the time period near such minima can be evaluated
using the small oscillation approximation \cite{smerzi97}, which
yields
\begin{equation}
\tau_{\text{SO}}=  \frac{\pi \hbar }{ \mathcal{ K}_i  \sqrt{   \Lambda_i+1 }}  \,,
\label{tauso}
\end{equation}
for $i= 0;1$ with  $ \mathcal{ K}_0= | \Re(K)|$ and  $ \mathcal{K}_1= | \Im(K)|$.

\subsubsection{ Closed orbits involving  $\pi$ and plasma modes }

For $\pi$ and plasma modes one has the following vortex coordinates,
\begin{equation}
Y_v^{(1)} (t)= \left( \frac{\varphi(t)}{\pi} + \, 2l_y+1  \right) \, \frac{  \pi \hbar}{ m q_0 \Omega} \,,
\label{vortyreal}
\end{equation}
\begin{equation}
X_v^{(0)} (t)= \left( -\frac{\varphi(t)}{\pi} + \, 2l_x+1  \right) \, \frac{  \pi \hbar}{ m q_0 \Omega} \,,
\label{vortxreal}
\end{equation}

\begin{figure}[!h]
 \includegraphics[width= 1\columnwidth]{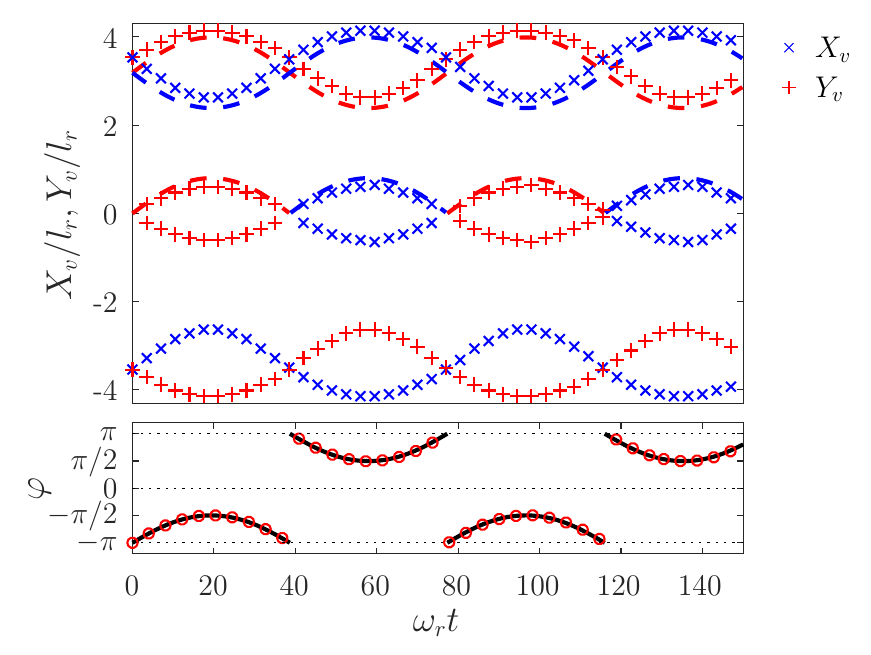}%
 \caption{\label{fig7} Time evolution of the vortices coordinates (top
   panel) as depicted in Fig. \ref{figom15} for
   $\Omega/ 2 \pi= 35$Hz. The dashed red and blue lines correspond to
   Eqs (\ref{vortyreal}) and (\ref{vortxreal}), respectively.  The
   corresponding phase difference $\varphi(t)$ is depicted in the
   bottom panel.  The initial conditions correspond to
   $\varphi(0)=\pi $, $N_0(0)= 2530 $, and $N_1(0)= 2470$.  The dashed
   red and blue lines represent the predicted values from
   Eqs. (\ref{vortyreal}) and (\ref{vortxreal}), respectively, with
   $ l_y,l_x \in \{0;1\} $. In the bottom panel, the phase differences
   extracted from GP simulations and the MM model is depicted with red
   circles and a solid black line, respectively. }
\end{figure}

In the case $ \Re(K) < 0$, $\pi$-modes appear in the phase-space
diagram.  Such orbits give rise to closed orbits of vortices.  Each
turning point of $\varphi$ is related to a turning point of the vortex
coordinate along the low-density path. The other coordinate in the
transversal direction moves from one site border to the neighboring
one due to the sign change of $Z$.  In Fig. \ref{fig7} we show
$Y_v^{(k)}(t)$ for $ \Delta N= 60$ and $\Omega / 2 \pi =35$Hz,
corresponding to a negative value
$ \Re(K)= -4.07 \times 10^{-4}\hbar \omega_r$.  The corresponding
critical particle difference yields $ \Delta N_c= 84.7$ and the small
oscillation period $ \tau_{\text{SO}}= 65.4 \omega_r^{-1}$.

If such vortex orbits are centered in a positive coordinate, their
periods coincide with the period of the orbit in the phase space, and
the shape of the trajectory is ellipsoidal. In Fig. \ref{fig7}, such
is the case of the vortex oscillation around
$Y_v^{(1)}= X_v^{(0)} \simeq 3.2 l_r$.  On the other hand, there
exists two vortices that move around $x=y=0$ with trajectories that
encircle such a point and are always nearer the site with a lower
number of particles. Such vortices pass from a $(k)$ junction with
$Y_v^{(k)}$ ( $X_v^{(k)}$) to a $(k+1)$ junction with $X_v^{(k+1)}$
($Y_v^{(k+1)}$), when $\varphi= \pi$ and have turning points in the
same low-density paths at the turning points of $\varphi (t)$ crossing
to the neighborhood of the sites of lower number of particles. Such
vortices display a ``flower''-type orbit.  As a consequence, the closed
orbit of such vortices has twice the period of $\varphi(t)$. Moreover,
since $\Delta N(0)=60$ is near the critical imbalance, the small
oscillation approximation does not apply here, as it may be seen from
the bottom panel of Fig. \ref{fig7}.

\begin{figure}[!h]
 \includegraphics[width=0.9\columnwidth]{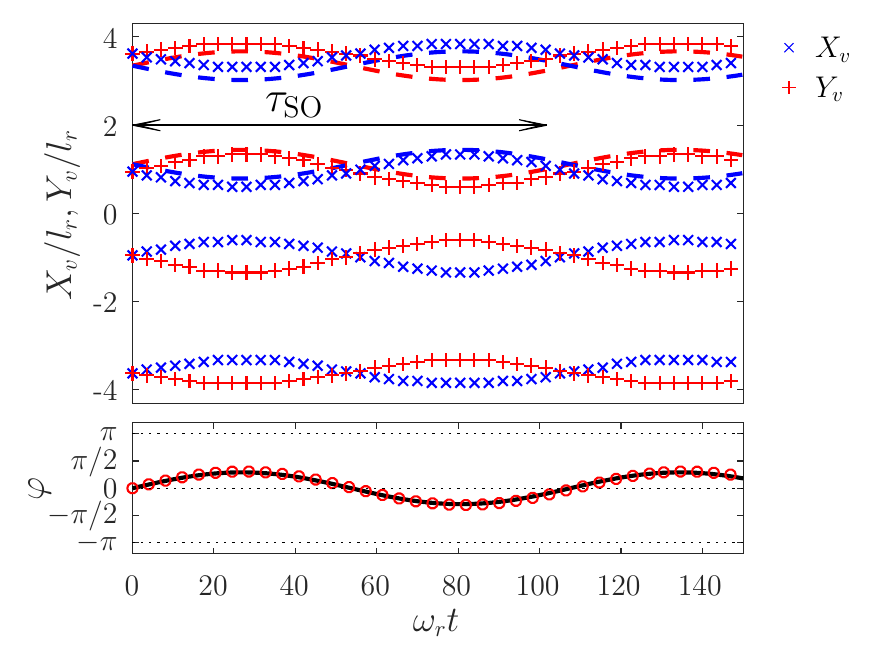}%
 \caption{\label{fig8} Same as Fig. \ref{fig7} for a rotation
   frequency $\Omega/2 \pi= 50$Hz , and for the initial conditions
   $\varphi(0)=0 $, $N_0(0)= 2512.5$, and $N_1(0)=2487.5$.  The time
   period $\tau_{\text{SO}}= 101.8 \omega_r^{-1}$ is indicated with an
   horizontal arrow. The indices $l_x$ and $l_y$ of the estimates from
   Eqs. (\ref{vortyreal}) and (\ref{vortxreal}), take the values
   $ \{0;1\} $.  }
\end{figure}

For $ \Re(K) > 0$ plasma modes are present and the central
``flower''-type-vortex dynamics is absent. For low frequencies, the
estimate of the stationary position yields a value outside the
lattice, and then, only part of the orbits may appear.  In Fig.
\ref{fig8}, we show such dynamics with a large rotation frequency
$\Omega/(2 \pi)= 50 $Hz which corresponds to a hopping value
$ \Re(K) = 1.68 \times 10^{-4}\hbar \omega_r $ and a critical particle
difference $ \Delta N_c = N_0-N_1= 54.4$. In this case, two vortices
move along each semiaxis around their corresponding stationary
position, and the small oscillation approximation yields
$\tau_{\text{SO}}= 101.8 \omega_r^{-1}$, which is in good accordance
with the period of the vortex orbits.

\subsubsection{ Vortex dynamics for  symmetric initial conditions with $ \pm \pi/2$- modes}

In this section, we analyze the vortex dynamics for which the initial
populations and phase differences verify $n_k(0)=n_{k+2}(0) $ and
$\varphi_k(0) + \varphi_{k+1}(0) = \pi$.  Then, the vortex coordinates
yield
\begin{equation}
Y_v^{(1)} (t)= \left( \frac{\varphi(t)}{\pi} + \, 2l_y+1  \right) \, \frac{  \pi \hbar}{ m q_0 \Omega} \,,
\label{vortyimag}
\end{equation}
and
\begin{equation}
X_v^{(0)} (t)= \left( -\frac{\varphi(t)}{\pi} + \, 2  l_x  \right) \, \frac{  \pi \hbar}{ m q_0 \Omega}\,.
\label{vortximag}
\end{equation}
Depending on the selected orbit, the central vortex can invert its
charge during the dynamics.  For low frequencies where
$ \Im(K) \le 0$, only one vortex can appear along each semiaxis near
the lattice border; hence we will focus on the case $ \Im(K) > 0$
given that a more interesting vortex dynamics emerges. In particular,
we will consider the rotation frequency $\Omega/2 \pi= 45$Hz, for
which the imaginary part of the hopping parameter yields
$ \Im(K) = 3.4138 \times 10^{-4} \hbar \omega_r $ and
$ \Delta N_c= 77.48$.

\begin{figure}[!h]
  \includegraphics[width= 1 \columnwidth]{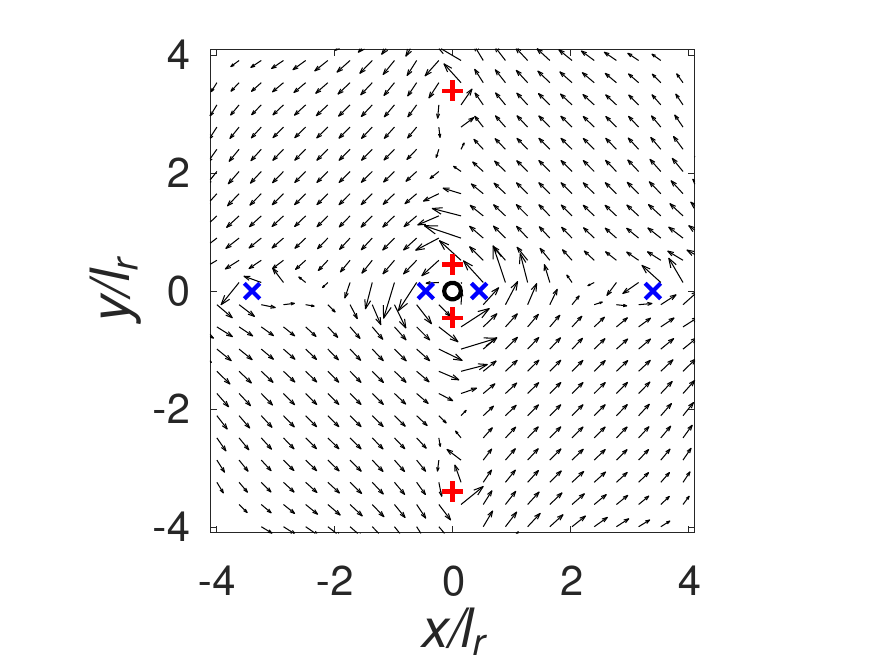}%
  \caption{\label{figsta} Stationary vortex configuration in the
    $(x,y)$- plane, corresponding to the state $\psi_{-1}(\mathbf{r})$
    obtained from Eq. (\ref{GProtstatic}), for $\Omega/ 2 \pi=
    45$Hz. In connection to the TM model, such a state is related to
    the point $\varphi = - \pi /2 $ and $Z=0$ of the phase-space
    diagram.  The vortices in the $ y$ and $x$ axes are marked with
    red plus sign symbols and blue crosses, respectively. The central
    antivortex is shown as a black-hollow circle. The black arrows
    represent the velocity field.  }
\end{figure}

In Fig. \ref{figsta} we show the vortex stationary configuration for
the GP state $\psi_{-1}(\mathbf{r})$ obtained by solving
Eq. (\ref{GProtstatic}) that corresponds to the Hamiltonian minimum
which is located at $Z=0$ and $\varphi=-\pi/2$, as can be verified
from Eq. (\ref{statn}).  Such vortex positions can be estimated using
Eqs. (\ref {vortyimag}) and (\ref{vortximag}) using
$ \varphi(t) =-\pi/2$, which yield
$Y_v^{(1)} = X_v^{(0)} \simeq 0.6 l_r$ and $ 3.1 l_r$, for $l_y= 0$
and $l_y= 1$, respectively. We note that
$\psi_{-1}(\mathbf{r}) = \psi_{\text{MM}}(\mathbf{r})$ given by
Eq. (\ref{orderparameter}) with $n_k=0.25$, $\phi_{-1}= \pi/2 $,
$\phi_0= 0$, $\phi_1= - \pi/2 $, and $\phi_2= \pi $.

\begin{figure}[!h]
 \includegraphics[width= 1.1 \columnwidth]{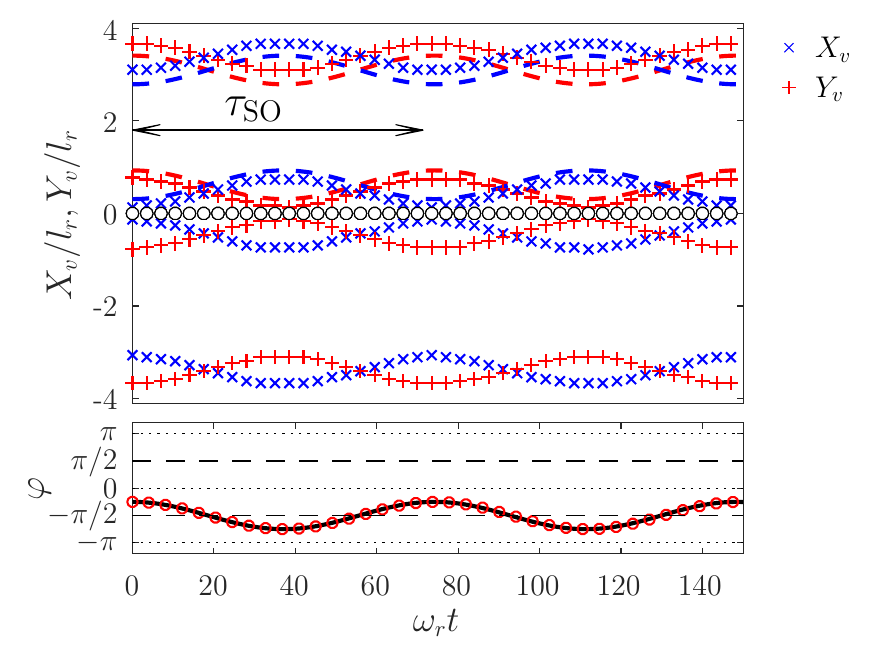}
 \caption{\label{fig12} Vortex coordinates as functions of time (top
   panel) extracted from the MM order parameter using the plaquette
   method, for $\Omega/ 2 \pi= 45$Hz.  The blue crosses and red plus
   signs of the positive coordinate values indicate $ X^{(0)}_v(t)$
   and $Y^{(1)}_v(t)$, respectively, whereas the same symbols are
   utilized for the negative values which correspond to
   $ X^{(2)}_v(t)$ and $Y^{(-1)}_v(t)$.  The black circles indicate
   the central antivortex.  The dashed red and blue lines represent
   the predicted values from Eqs. (\ref{vortyimag}) and
   (\ref{vortximag}), respectively.  The initial conditions correspond
   to $\varphi(0)= - \pi /4 $ and $Z=0$.  The time period
   $\tau_{\text{SO}}=71.29 \omega_r^{-1}$ is represented by an
   horizontal arrow.  In the bottom panel, the phase difference is
   shown as a function of time, depicted with red circles and a solid
   black line, for the GP and MM model results, respectively.  }
\end{figure}

Small phase-difference oscillations around $- \pi/2$ may be viewed in
the bottom panel of Fig. \ref{fig12} with the corresponding vortex
dynamics in top panel.  There, the vortices oscillate but do not reach
the origin of the $x-y$ plane.  Hence, the central antivortex located
along the $z$-axis never change its charge. The small oscillation
period obtained through Eq. (\ref{tauso}) yields
$\tau_{\text{SO}}=71.29 \omega_r^{-1}$, which is in good accordance
with the numerical results. The selected orbit in the phase space
diagram is depicted in the bottom panel of Fig. \ref{diagrama15}
which, as it can be seen, is entirely inside the dashed area in
accordance with the permanence of the central antivortex.

\begin{figure}[!h]
 \includegraphics[width= 1.1 \columnwidth]{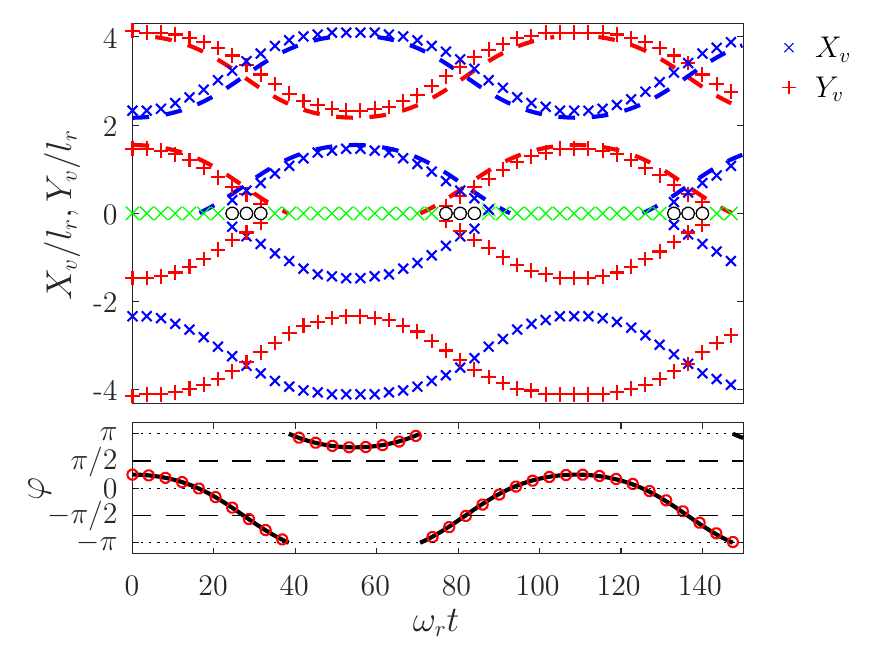}%
 \caption{\label{fig14}Vortex coordinates as functions of time (top
   panel) extracted from the MM order parameter and depicted as in
   Fig. \ref{fig12}.  The green crosses and the black circles indicate
   the central vortices and antivortices, respectively.  The dashed
   red and blue lines represent the predicted values from
   Eqs. (\ref{vortyimag}) and (\ref{vortximag}), respectively, with
   $ l_y=0,1$ and $l_x=0,1$.  The initial conditions correspond to
   $\varphi(0)= \pi /4 $ and $Z=0$, and the rotation frequency to
   $\Omega/ 2 \pi= 45$Hz.  In the bottom panel the phase difference
   $\varphi(t)$ is depicted as in Fig. \ref{fig12} }
\end{figure}

Phase-difference oscillations with a larger amplitude can be viewed in
Fig.  \ref{fig14}, where in this case the central vortex changes its
charge. Initially a central vortex is present, whereas when two
vortices depart from the origin along the $x$-axis, depicted with the
blue crosses in such a graph, a vortex-antivortex pair should be
created in order to provide one of the vortices that depart as well as
the central antivortex.  An instant later, the other two vortices
coming along the $y$-axis, which are indicated with red plus signs,
provoke the annihilation of a vortex-antivortex pair and hence a
single central vortex survives.

\subsection{ Open orbits related to self-trapping regime}

In the self-trapping regime, the imbalance oscillates around a
nonvanishing value and the phase difference exhibits a monotonously
increasing, or decreasing, behavior that is referred to as a running
phase. The oscillation period for the Hamiltonian $H_i$ can be
approximated by \cite{nig17}
\begin{equation}
\tau_{\text{st}}=  \frac{ Z_0 \pi \hbar }{ 2 \mathcal{K}_i } \left[1- \sqrt{ 1-  \frac{4}{ \Lambda_i  Z_0^2 }} \, \right]  \,,
\label{taustre}
\end{equation}
where $Z_0$ is the imbalance value at the maximum of each
corresponding open orbit.

\begin{figure}[!h]
 \includegraphics[width=1 \columnwidth]{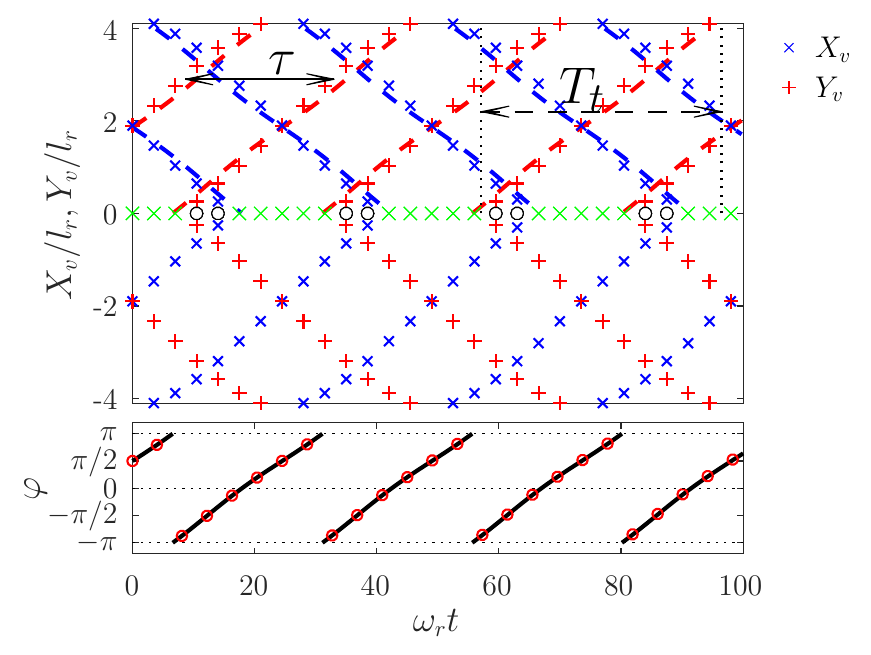}%
 \caption{\label{fig16} Vortex positions and phase difference as
   functions of time are shown in the top and bottom panels
   respectively and depicted as in Fig. \ref{figom30}. The initial
   conditions are $\varphi(0)= 0.5 \pi $, $N_0(0)= 2550 $, and
   $N_1(0)=2450$. The horizontal arrows indicate the period
   $\tau_{\text{st}}= 24.4 \omega_r^{-1}$ and the traveling time
   $T_t=39.4 \omega_r^{-1}$.  The rotation frequency corresponds to
   $\Omega/ 2 \pi= 45$Hz.  }
\end{figure}

An example of such types of evolutions is shown in Fig. \ref{fig16},
where in the bottom panel we depict the running phase difference for
the Hamiltonian $H_1$ and the related vortex dynamics is shown in the
top panel.  The rotation frequency corresponds to $\Omega/2\pi = 45$
Hz.  From the bottom panel of Fig. \ref{diagrama15} it can be seen
that the maximum value of $Z$ is achieved at $\varphi=- \pi/2$, which
for the initial conditions considered in the present evolution, yields
$Z_0\simeq 0.0253$. Then, from Eq. (\ref{taustre}) one obtains an
estimate of the time period $\tau_{\text{st}}= 24.4 \omega_r$ which is
in good agreement with the data of Fig. \ref{fig16}.  Given that
$\dot\varphi(t)$ is always nonzero, the vortex trajectories do not
possess turning points. If $N_0>N_1$ the vortices enter by the
$x$-axis and depart from the system along the $y$-axis. Since the
instant when vortices arrive at the $z$- axis do not coincide with the
instant they leave it, there exist an inversion of the central vortex
charge.  The time such vortices spend in moving from the trap center
up to the end of the lattice can be approximated by,
\begin{equation}
T_t=  \frac{ L m q_0   \Omega \tau_{\text{st}}  }{  \hbar  2 \pi}  \,,
\label{timetst}
\end{equation}
where $L=4 l_r$ denotes the largest absolute value of a vortex
coordinate inside the lattice.  Then, for the data of
Fig. \ref{fig16}, we obtain $T_t=41.4 \omega_r^{-1}$ (dashed arrow) in
good agreement with the numerical result.

\subsection{ Comparison between the MM model and GP results for  the position of the vortices}

\begin{figure}[!h]
  \includegraphics[width=\columnwidth]{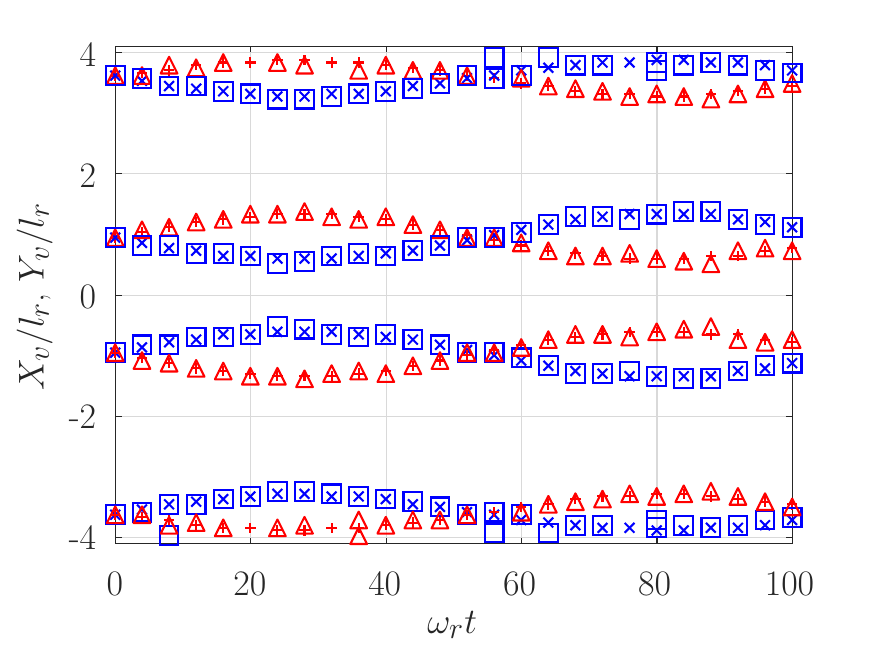}
  \caption{\label{fig17} Comparison between the GP and MM model vortex
    position as a function of time for the same conditions of
    Fig. \ref{fig8}.  The blue squares (crosses) and red
    triangles (plus signs) indicate the coordinates of the
    vortices extracted from $\psi_{\text{GP}} ({\mathbf r},t)$ (
    $\psi_{\text{MM}} ({\mathbf r},t)$) order parameter along the
    $x$ and $y$ axes, respectively. }
\end{figure}

\begin{figure}[!h]
  \includegraphics[width= 1.1 \columnwidth]{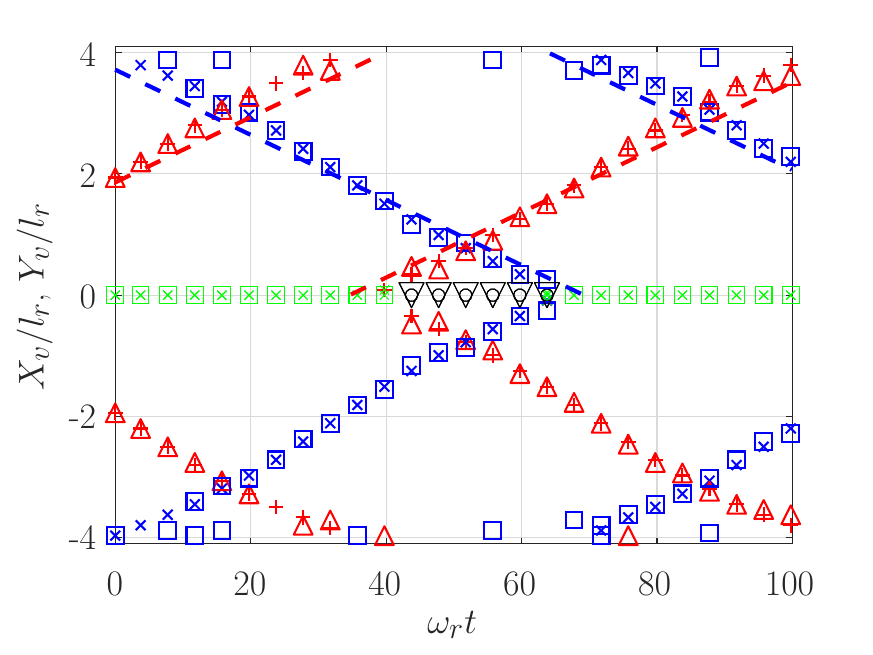}
  \caption{\label{fig18} Comparison between the GP and MM results
  as function of time, for $N_0=2520$, $N_1=2480$, $\varphi(0)=0$
  and $\Omega/2 \pi=30$Hz.  The red and blue symbols are the same
  of Fig. \ref{fig17}, whereas the green squares (crosses) and
  black triangles (circles) correspond to the central vortices
  and antivortices, respectively, extracted from
  $\psi_{\text{GP}} ({\mathbf r},t)$ ($\psi_{\text{MM}}
  ({\mathbf r},t)$) order parameter. }
\end{figure}

It may be seen from the bottom panels of the previous figures that in
all cases the MM model accurately reproduces the GP phase difference,
which constitutes the main ingredient for determining the vortex
position.  Nevertheless, small deviations in the vortex location can
occur due to density fluctuations of the time-dependent order
parameter $\psi_{\text{GP}} ({\mathbf r},t)$.  Such small effects can
be viewed in Fig. \ref{fig17}, where we have depicted the vortex
coordinates extracted from $\psi_{\text{GP}} ({\mathbf r},t)$ using
the plaquette method.  The GP results are depicted in
Fig. \ref{fig17}, with squares and triangles for the same conditions
as in Fig. \ref{fig8}.  Minor deviations between the squares and
crosses or between the triangles and plus symbols can be observed
except at the border of the condensate, where the density is very low
and some larger fluctuations are present.  During the evolution,
short-living vortex-antivortex pairs can also be generated, and hence
they slightly modify the position of the more stable vortices which we
are interested in.  For completeness, in Fig. \ref{fig18} we also
include a case where central vortices are present and one can observe
a good accordance of the MM model with GP results, and both are well
predicted by the estimate too.

\section{\label{sec:conclu}Conclusions}
We have shown that the dynamics of vortices in a rotating lattice is
determined by the phase differences between sites rather than by
density gradients as it occurs in nonrotating systems. In particular,
we demonstrate that different vortex dynamics can be engineered via
the implementation of appropriate initial conditions on populations
and phases in a rotating ring lattice.  Such dynamics include open
orbits associated with a running phase difference, in which case the
vortices enter the lattice along one axis and exit along the other one
without exhibiting turning points. The time of permanence of the
vortex inside the lattice can be estimated by using only the on-site
interaction parameter in some cases, or evaluated via formulas
developed for the self-trapping regime of a TM model. For particular
initial phase differences, the dynamics involve the creation and
annihilation of vortex-antivortex pairs at the center of the trap. The
phase-space diagram of the associated Hamiltonian can be partitioned
into regions where the central vortex or antivortex prevails, which
hence allows us to predict when such creation or annihilation
processes occur. Closed orbits of vortices may be generated by
choosing initial conditions near the minimum of the TM-model
Hamiltonian. The related period turns out to be the same as that of
the orbits of the model and can be easily evaluated in the case of
small oscillations. The points on the $x-y$ plane around which such
oscillations occur approach the origin as the rotation frequency
increases, and then more vortices oscillate around different
points. In general, for open orbits, the velocity of the vortices
inside the lattice decreases when the rotation frequency increases. We end by
stressing that the accuracy of our approach is independent of the specific details
of the trapping potential and on-site populations. Such an approach 
only requires that the system is formed by weakly linked spheroidal condensates
with  all  their  principal axes  parallel  to the rotation axis.  
In summary, we find that the present system constitutes a
useful setting for studying different vortex processes, where both the orbits
and the time periods can be predicted and applied in different settings.

\newpage

\appendix*

\section{Equations of motion and parameters of the multimode model}

The equations of motion for $n_k$ and $\varphi_k$ can be obtained by
inserting the MM model order parameter of Eq.~(\ref{orderparameter})
into Eq.~(\ref{gp}), yielding \cite{rot20}
\begin{widetext}
  \begin{align}
 \hbar\,\frac{dn_k}{dt}& =  2 |J| \left[ \sqrt{n_k \, n_{k+1}} \, \sin (\varphi_{k+1}+\theta _J) 
- \sqrt{n_k \, n_{k-1} } \, \sin (\varphi_k+\theta _J) \right ]\nonumber\\
&-  2 |F| \left[ \sqrt{n_k \, n_{k+1} } (n_k + n_{k+1} ) \, \sin (\varphi_{k+1}+\theta _F) \right.- 
\left.\sqrt{n_k \, n_{k-1} } (n_k + n_{k-1} ) \, \sin (\varphi_k+\theta _F)\right] \, ,
\label{ncmode1hn}\\
 \hbar\,\frac{d\varphi_k}{dt} & =    ( n_{k-1} -n_{k}) N  U_{\text{\text{eff}}}    
- \alpha (  n_{k-1} - n_{k}) N U \left[   N_s (n_{k-1}+  n_{k})-2 \right]  \nonumber\\
&+  |J| \left[ \left(\sqrt{\frac{n_k}{ n_{k-1}}} 
- \sqrt{\frac{n_{k-1} }{ n_k}}\,\right) \, \cos (\varphi_k+\theta _J) \right.+
 \left. \sqrt{\frac{n_{k-2}}{ n_{k-1} }} \, \cos (\varphi_{k-1}+\theta _J) 
- \sqrt{\frac{n_{k+1} }{ n_k}} \, \cos (\varphi_{k+1}+\theta _J) \right]\nonumber\\
&-  |F|  \left[ \left( n_k \sqrt{\frac{n_k}{ n_{k-1} }} - n_{k-1} \sqrt{\frac{n_{k-1}}{ n_k}}\,\right)
 \, \cos (\varphi_k+\theta _F)   \right.\nonumber\\
  &+ \left( 3\, \sqrt{n_{k-2} \, n_{k-1}} + n_{k-2} \sqrt{\frac{n_{k-2}}{ n_{k-1}}}\,\right) 
 \, \cos (\varphi_{k-1}+\theta _F)  
 \left.- \left(  3\, \sqrt{n_{k+1} \, n_k} + n_{k+1} \sqrt{\frac{n_{k+1}} { n_k}}\,\right)  \, 
\cos (\varphi_{k+1}+\theta _F)\right], \nonumber \\
\label{ncmode2hn}
\end{align}
\end{widetext}
where $\varphi_k=\phi_k-\phi_{k-1}$.  The standard hopping parameter
$J$ and the interaction-driven one $F$ are given by
\begin{eqnarray}
&& J = -\int d^3r\, w_{0}^*(\hat{H}_0-\Omega \hat{L}_z)w_{1}, \label{Jjk} \\
&& F =-gN \int d^3r\, (w_{0}^*)^2 w_{0} w_{1}. \label{R} \\
\nonumber
\end{eqnarray}

For $\Omega\neq 0$, such parameters $J$ and $F$ become complex numbers
with phases $\theta_J + \pi$ and $\theta_F$, respectively. On the
other hand, as described in Ref. \cite{jezek13a}, the effective
interaction parameter $U_{\text{eff}}$ is obtained by calculating the
on-site interaction energy parameter
\begin{equation}
 U = g \int d^3r\, | w_{0}|^4,
\label{eq:Ueff}
\end{equation}
and evaluating the parameter $\alpha$, which takes into account the
variation of the on-site interaction energy with the occupation
number. Then, the effective interaction parameter yields
$U_{\text{\text{eff}}}=(1-\alpha)U$.

For our system setup in the nonrotating case, the standard hopping
yields $ J=-6.60\times 10^{-4} \hbar \omega_r $, the
interaction-driven hopping parameter
$F= 2.08\times 10^{-3} \hbar\omega_r$, the on-site interaction energy
$U= 3.16\times 10^{-3} \hbar\omega_r$, and the effective on-site
interaction energy $U_{\text{eff}}=2.269 \times10^{-3}\hbar\omega_r$,
being $ 1-\alpha \simeq 0.719 $. For more details, see
Ref. \cite{mauro4p}.

For nonvanishing $\Omega$ values, in the present work, the equations
of motion will involve only a single hopping $K$ defined by
$K=2J+F= |K| e^{i \theta_K}$.  From the definitions of the hopping
parameters, it is easy to demonstrate that the real and imaginary
parts of the hopping $K$ are related to the stationary states energies
$E_i$ by,
\begin{equation}
 2 \Re({K})=  E_2 -E_0
\end{equation}
and
\begin{equation}
 2 \Im({K})=  E_1 -E_{-1}.
\end{equation}
With increasing values of $\Omega$, the absolute value of such hopping
parameter monotonously decreases; whereas, given that the relative
values of the energies change, both the real and imaginary parts of
$K$ alternate signs as functions of $\Omega$.

As a consequence of the geometry of the trap, when the system is
subject to rotation the on-site localized states acquire a phase
gradient which gives rise to a related, almost homogeneous, velocity
field \cite{rot20}.  Such a particular phase profile determines the
arguments of the hopping parameters, which turn out to be equal to a
single phase $\Theta \equiv \theta_K= \theta_J =\theta_F$ that obeys
\begin{equation}
\Theta = -   \frac{ m }{\hbar} \Omega r^2_{\text{cm}} \sin{(2 \pi/N_s)} \, ,
\label{teta}
\end{equation}
where $r_{\text{cm}}$   can be roughly  approximated in terms of the intersite parameter.

Finally, with respect to the interaction energy parameters $U$ and
$U_{\text{eff}}$, they remain almost constant as functions of $\Omega$
\cite{rot20}.  It is worthwhile to mention that, even though in the
Thomas-Fermi (TF) approximation, the effective interaction energy
parameter is reduced with respect to $U$ by a factor of 7/10, 3/4, or
5/6, for either three, two, or one dimensions, respectively. Here, the
factor $ 1-\alpha\simeq 0.719 $ has been obtained numerically, as
described in \cite{rot20}, which yields a slightly higher value than
that corresponding to a TF 3D system.


\begin{thebibliography}{99}
%
\bibitem{don91}
R. J. Donnelly, Quantized Vortices in Helium II
(Cambridge University Press, Cambridge, 1991).
%
\bibitem{dal99} F.  Dalfovo, S. Giorgini, L. P. Pitaevskii,  and  S. Stringari,  Rev. Mod. Phys.  \textbf{71}, 483 (1999).
%
\bibitem{fet09}
 A. L. Fetter,  Rev. Mod. Phys. \textbf{81}, 647  (2009).
%
\bibitem{mat99}
 M. R. Matthews, B. P. Anderson, P. C. Haljan, D.S. Hall,
C. E. Wieman, and E. A. Cornell, Phys. Rev. Lett. \textbf{83}, 2498
(1999).
%
\bibitem{brian00}
 B. P. Anderson, P. C. Haljan, 
C. E. Wieman, and E. A. Cornell, Phys. Rev. Lett. \textbf{85}, 2857
(2000).
%
\bibitem{mad00}
 K. W. Madison, F. Chevy, W. Wohlleben, and J. Dalibard,
Phys. Rev. Lett. \textbf{84}, 806 (2000).
%
\bibitem{abo01}
 J. R. Abo-Shaeer, C. Raman, J. M. Vogels, and W. Ketterle,
 Science \textbf{292}, 476 (2001).
%
\bibitem{enge03}
 P.  Engels, I. Coddington, P. C. Haljan, V. Schweikhard, and
E. A. Cornell, Phys. Rev. Lett.\textbf{90}, 170405 (2003).
%
\bibitem{bre04}
 V. Bretin, S. Stock, Y. Seurin, and J. Dalibard, Phys. Rev. Lett.
\textbf{92}, 050403 (2004).
%
\bibitem{sto05}
 S. Stock, B. Battelier, V. Bretin, Z. Hadzibabic, and J. Dalibard,
Laser Phys. Lett. \textbf{2}, 275 (2005).
%
\bibitem{ryu07}
 C. Ryu, M. F. Andersen, P. Cladé, V. Natarajan, K. Helmerson,
and W. D. Phillips, Phys. Rev. Lett. \textbf{99}, 260401 (2007).
%
\bibitem{ec14}
 S. Eckel, J. G. Lee, F. Jendrzejewski, N. Murray, C. W. Clark,
C. J. Lobb, W. D. Phillips, M. Edwards, and G. K. Campbell,
Nature (London) \textbf{506}, 200 (2014).
%
\bibitem{wil22}
 K. E.Wilson, E. C. Samson, Z. L. Newman, and B. P. Anderson,
Phys. Rev. A \textbf{106}, 033319 (2022).
%
\bibitem{tun06}
S. Tung, V. Schweikhard, and E. A. Cornell, Phys. Rev. Lett.
\textbf{97}, 240402 (2006).
%
\bibitem{wi10}
R. A. Williams, S. Al-Assam, and C. J. Foot, Phys. Rev. Lett.
\textbf{104}, 050404 (2010).

\bibitem{fet05}
 A. L. Fetter,  B. Jackson, S. Stringari, Phys. Rev. A \textbf{71}, 
013605 (2005).
%
\bibitem{kim05}
 J. K. Kim, A. L. Fetter, Phys. Rev. A \textbf{72}, 023619 (2005).
%
\bibitem{jez08}
D. M. Jezek, P. Capuzzi,
M. Guilleumas, and R. Mayol,
Phys. Rev. A \textbf {78}, 053616  (2008).
%
%
\bibitem{sheehy04}
 D. E. Sheehy and L. Radzihovsky, Phys. Rev. A \textbf{70}, 
063620 (2004).
%
\bibitem{nil06}
 H. M. Nilsen, G. Baym, and C. J. Pethick, Proc. Natl.
Acad. Sci. USA \textbf{103}, 7978 (2006).
%
\bibitem{jez108}
 D. M. Jezek and H. M. Cataldo, Phys. Rev. A \textbf{77}, 043602
(2008).
%
%
%
\bibitem{gros18}
A. J. Groszek, D. M. Paganin, K. Helmerson, and T.  P. Simula,
 Phys. Rev. A \textbf{97}, 023617 (2018).
%
\bibitem{nav13}
R. Navarro, R. Carretero-González, P. J. Torres, P. G. Kevrekidis,
D. J. Frantzeskakis, M. W. Ray, E.  Altuntas¸, and D. S. Hall,
 Phys. Rev. Lett.  \textbf{110}, 225301 (2013).
%
\bibitem{zhang19}
T.  Zhang, J. Schloss, A. Thomasen,  L. J. O’Riordan,
  T.  Busch, and A. White,  Phys. Rev. Fluids \textbf{4}, 054701 (2019).
%
\bibitem{frei10}
D. V.  Freilich, D. M. Bianchi, A. M. Kaufman, T. K. Langin, and D. S. Hall,
   \textbf{329}, 1182 (2010).
%
\bibitem{nee10}
T. W. Neely,  E. C. Samson,  A. S. Bradley,  M. J. Davis,  and  B. P. Anderson,
 Phys. Rev. Lett.  \textbf{104}, 160401 (2010).
%
\bibitem{nee13}
T.  W.  Neely,  A. S. Bradley,  E. C. Samson,  S. J. Rooney,  E. M. Wright,  K. J. H.
Law,  R. Carretero-González,  P. G. Kevrekidis,   M. J. Davis,  and  B. P. Anderson,
 Phys. Rev. Lett.  \textbf{111}, 235301 (2013).
%
\bibitem{sera15}
S. Serafini, M. Barbiero, M. Debortoli, S. Donadello,
F. Larcher, F. Dalfovo, G. Lamporesi,  and G. Ferrari,
 Phys. Rev. Lett.  \textbf{115}, 170402 (2015).
%
\bibitem{ma06}
A.  M.  Mateo and V. Delgado
 Phys. Rev. Lett.  \textbf{97}, 180409 (2006).
%
%
\bibitem{se10}
J. A. Seman,  E. A. L. Henn, M. Haque, R. F. Shiozaki, E. R. F.  Ramos, M. Caracanhas, P. Castilho,
C. Castelo Branco, P. E. S. Tavares, F. J. Poveda C., G. Roati, K. M. F.   Magalh\~aes, and V. S. Bagnato,
 Phys. Rev. A \textbf{82}, 033616
(2010).
%
%
%
%
\bibitem{abad15}
M. Abad, M. Guilleumas, R. Mayol, F. Piazza, D. M. Jezek, and A. Smerzi,
EPL,  \textbf{109}, 40005 (2015).
%
\bibitem{don14}
Donadello, S. Serafini, M. Tylutki, L. P. Pitaevskii, F. Dalfovo,
G. Lamporesi, and G. Ferrari, Phys. Rev. Lett.  \textbf{113}, 065302
(2014).
%
\bibitem{jez16}
 D. M. Jezek, P. Capuzzi, H. M. Cataldo, Phys. Rev. A \textbf{93},
023601 (2016).
%
\bibitem{rot20}
M. Nigro, P. Capuzzi, and D. M. Jezek, J. Phys. B: At. Mol. Opt. Phys. \textbf{53}, 025301 (2020).
%
\bibitem{je23}
D. M. Jezek  and  P. Capuzzi,
Phys. Rev. A \textbf {108}, 023310  (2023).
%

\bibitem{Poli2023}
  E. Poli,  T. Bland, S. J. M. White,
  M. J. Mark, F. Ferlaino, S. Trabucco, and M.
  Mannarelli, Phys. Rev. Lett. \textbf{131}, 223401 (2023).
\bibitem{Bland2024} T. Bland, F. Ferlaino, M.
  Mannarelli, E. Poli, and S. Trabucco, Few-Body Systems
  \textbf{65}, 81 (2024).
  
\bibitem{mauro4p} M. Nigro, P. Capuzzi, H. M. Cataldo, and
D. M. Jezek, Phys. Rev. A
 \textbf{ 97}, 013626 (2018).
%
\bibitem{cat11}
H. M. Cataldo and D. M. Jezek, Phys. Rev. A \textbf{ 84},  013602 (2011).
%
\bibitem{albiez2005} M. Albiez, R. Gati, J. Fölling, S. Hunsmann,
  M. Cristiani, and M. K. Oberthaler, Phys. Rev. Lett. \textbf{95},
  010402 (2005).
  %
\bibitem{nig22}
M. Nigro, P. Capuzzi, and D. M. Jezek,  Eur. Phys. J. D \textbf {76}, 12 (2022).
%
\bibitem{foster10}
C. J. Foster, P. B. Blakie,  and  M. J. Davis, Phys. Rev. A \textbf{ 81}, 023623 (2010).
%
%
%
\bibitem{smerzi97} A. Smerzi, S. Fantoni, S. Giovanazzi, and S. R. Shenoy, 
Phys. Rev. Lett. \textbf{79}, 4950 (1997).
%
\bibitem{nig17}
M. Nigro, P. Capuzzi, H. M. Cataldo, and D. M. Jezek,  Eur. Phys. J. D \textbf {71}, 297 (2017).
%
%
\bibitem{jezek13a} D. M. Jezek, P. Capuzzi,  and H. M. Cataldo, Phys. Rev. A
 \textbf {87}, 053625 (2013).
%

\end{thebibliography}
\end{document}